\documentclass[amsmath,amssymb,reprint,letterpaper,aps,superscriptaddress]{revtex4-2}
\usepackage{graphicx,xcolor} 
\usepackage{bm}
\usepackage{textcomp}
\usepackage{gensymb}
\usepackage{array,multirow}
\newcolumntype{K}[1]{>{\centering\arraybackslash}m{#1}}
\usepackage{makecell}
\usepackage{tablefootnote}
\usepackage[english]{babel}
\usepackage{blindtext}
\usepackage{lipsum}
\usepackage{amssymb}
\usepackage{amsfonts}
\usepackage{amsmath}

\graphicspath{{./figures/}}

\begin{document}

\title{Encounter Times of Intermittently Running Particles}

\author{Lizzy Teryoshin}
\affiliation{Department of Physics, University of California, San Diego, La Jolla, CA 92093}	

\author{Mario Hidalgo-Soria}
\affiliation{Department of Physics, University of California, San Diego, La Jolla, CA 92093}	

\author{Elena F. Koslover}
\email{ekoslover@ucsd.edu}
\affiliation{Department of Physics, University of California, San Diego, La Jolla, CA 92093}
\date{\today}
\preprint{}

\begin{abstract}

Intracellular processes often rely on the timely encounter of mobile reaction
partners, including intermittently motor-driven organelles. The underlying
cytoskeletal network presents a complex landscape that both directs particle
movement and introduces quenched disorder through filament organization.
We investigate the mean first encounter times for pairs of intermittently
processive and diffusive particles, moving in two dimensions with and without a fixed filament network.
In unstructured domains, increasing particle run-length enhances exploration of the domain, but tends to slow down the encounter times compared to equivalent diffusing particles. Encounters for long-running particles occur preferentially near the periphery, contrasting with bulk encounters for the purely diffusive case.
When particles are unbiased in their runs along dense filament networks, encounters are shown to be well approximated by a continuum run-and-tumble model. For biased particles, regions of convergent filament orientation can serve as traps that slow the overall spatial exploration but can allow for faster encounter rates by funneling particles into regions of reduced dimensionality. These findings provide a framework for estimating intracellular encounter kinetics, highlighting the role of key physical features such as the effective diffusivity, run times, and network architecture.

\end{abstract}

\maketitle
\newpage

\section{Introduction}

The ability of intracellular particles to search out targets within the cell underlies a broad variety of biological processes~\cite{koslover2024searching}, including secretion~\cite{fourriere2020role}, signaling cascades~\cite{sorkin2009endocytosis,bakker2017egfr}, organelle maturation~\cite{york2023deterministic,cason2022spatiotemporal}, gene regulation~\cite{halford2004site,herbst2017regulated}, and self-assembly of organelles and protein complexes~\cite{devaux2010posttranscriptional,stehbens2012targeting}. Such search processes often rely on exploratory dynamics~\cite{kondev2025biological}, where randomly directed trajectories that lack a long-range guidance system must explore blindly until the target is successfully found. The lack of directional cues is particularly common in systems where the target itself is a mobile particle rather than a fixed cellular region. Examples include  early endosomes that need to encounter each other to proceed through the maturation pathway~\cite{york2023deterministic}, `social networks' of mitochondria whose encounters limit their ability to exchange material~\cite{chustecki2021network,holt2025diffusive}, and autophagosomes that must fuse with lysosomes to acidify their lumen and break down their cargo~\cite{cason2022spatiotemporal}.

Encounters and target search processes within the cell require partile mobility. Many subcellular components exhibit phases of diffusive transport, driven by active yet uncorrelated forces within the cytoplasm~\cite{lin2016active,drechsler2017active}. However, the cytoplasmic diffusivity is on the order of $10^{0}\mu\text{m}^2/\text{s}$ for large protein complexes~\cite{carlini2020microtubules,magiera2025measurement} and $10^{-3}\mu\text{m}^2/\text{s}$ for vesicles~\cite{lin2016active,koslover2016disentangling}, implying minutes to hours timescales for exploring typical-sized eukaryotic cells. To increase particle mobility, cells utilize active transport along cytoskeletal highways. Cargos recruit molecular motors such as kinesin, dynein, and various myosins, which walk processively along polarized microtubule or actin filaments~\cite{mogre2020getting}. Long-range motor-driven transport has been shown to play a role in endosome maturation~\cite{nielsen1999rab5,driskell2007dynein,flores-rodriguez2011roles}, mitochondrial distribution~\cite{misgeld2017mitostasis}, and the early secretory pathway in animal cells~\cite{presley1997er}.

Periods of active transport along cytoskeletal filaments are often interspersed with passive pauses, which may occur at microtubule intersections~\cite{balint2013correlative}, or may be caused by obstacle encounters, tug-of-war between oppositely directed motors, or dissociation from the cytoskeletal filament~\cite{mogre2020getting,hancock2014bidirectional,jongsma2023choreographing}. The run lengths between such pauses can be broadly distributed, ranging from a few hundred nanometers for lipid droplets~\cite{shubeita2008consequences} and dense core vesicles~\cite{ahmed2012mechanical} to over $10\mu$m for early endosomes~\cite{higuchi2014early}. Many cargos exhibit bidirectional motion along the filaments, occasionally reversing their direction when encountering intersections~\cite{ross2008kinesin} or when rebinding to a filament following a diffusive period~\cite{hancock2014bidirectional}. Others exhibit some amount of bias in preferentially walking towards the plus or minus ends of each filament~\cite{nielsen1999rab5,granger2014role}. The interplay of passive and active transport~\cite{mogre2018multimodal} as well as the directional bias of actively moving particles~\cite{hafner2018spatial,hafner2016spatial,ando2015cytoskeletal} can greatly alter both the spatial dispersion and the target search properties of intracellular components.

 A variety of modeling approaches have been used to describe the intermittent active transport motion typical of intracellular particles. These include creeper~\cite{mogre2018multimodal,campos2015optimal}, run-and-tumble~\cite{hafner2016run}, and random-velocity models~\cite{hafner2016spatial,hafner2018spatial,benichou2011intermittent,le_vot2022first,rupprecht2016optimal} that allow each particle to independently select the direction of each run, without the spatial correlations that might arise from an underlying network.
 Other models represent the cytoskeleton as an explicit network of filaments serving as spatially fixed highways for transport~\cite{ando2015cytoskeletal,hafner2016run,maelfeyt2019anomalous,mlynarczyk2019first,ciocanel2018modeling} .
 Typically, these models explore the search time to a fixed region of the domain, which could represent the outer cell
  boundary~\cite{ando2015cytoskeletal,maelfeyt2019anomalous}, a central target such as the nucleus~\cite{rupprecht2016optimal}, or a `narrow escape' through a small absorbing zone on the boundary~\cite{hafner2018spatial}. Run-and-tumble models have shown that intermediate values of run-length minimize the mean first passage time (MFPT) to a small target, since tumbles near the target increase the possibility of an encounter~\cite{rupprecht2016optimal,tejedor2011encounter}. Models of biased particles on explicit network structures have highlighted the importance of localized traps, which arise where microtubule ends of identical polarity converge together. Such traps can greatly slow down the first passage time across the domain, make target search times highly sensitive to the polarity of a few individual filaments, and lead to high variability in search times between different network realizations~\cite{ando2015cytoskeletal,mlynarczyk2019first}. Vortex-like traps are also observed in experimental measurements of particle trajectories on random actin filament networks reconstructed {\em in vitro}~\cite{scholz2016cycling}.
 
 More structured organization of the filament network can give rise to spatially inhomogeneous search processes, with distinct distributions of velocity vectors and switching kinetics in different subregions of the domain~\cite{schwarz2016optimality,hafner2020spatially}. For example, the growth of microtubules from a microtubule organizing center (MTOC) can result in preferentially radial cytoplasmic transport within the bulk of the cell, while short runs along a disordered actin meshwork yield unpolarized motion near the periphery. Such concentric regions with distinct transport properties have been shown to speed up search times for narrow peripheral targets~\cite{schwarz2016optimality,hafner2016spatial}. Increased filament density near the nucleus can also enhance the search rate to reach the peripheral boundary as a whole~\cite{ando2015cytoskeletal}.
 
 Fewer results are available for the influence of underlying network structure and kinetics on the rate of encounter between two particles which are both mobile. Encounters between identical diffusive particles generally occur more rapidly than when one particle is fixed, but the quantitative enhancement in the encounter time depends on the geometry of the domain and the location of the fixed particle~\cite{scott2021diffusive,scott2023endoplasmic,le_vot2022first}. For particles that engage in intermittent runs, their ability to find each other can be enhanced by imposing a radial bias that tends to make them converge towards the center, as might be expected for dynein-carried cargos on a microtubule network emanating from a single MTOC~\cite{hafner2018spatial}. However, in general it remains unclear how encounter times between mobile particles are modulated by changing run-length and by the explicit spatial organization of the filaments serving as transport highways.

In this work, we focus primarily on the encounter process between mobile particles that engage in intermittently persistent transport. We begin by considering run-and-tumble particles which select a new velocity direction upon each tumbling event. We show that increasing the length of active runs enhances the particles' ability to spread through the domain, but that the encounter times for run-and-tumble particles are higher than for diffusing particles with the same effective spreading rate. We then consider particles moving on explicit network structures, showing that for sufficiently dense networks and unbiased particles, the network structure has little effect on the encounter time. However, for particles that are biased towards one end of polarized filaments, network structures tend to give rise to traps. Such traps can speed up particle encounters by reducing the volume of space that particles have to search, but can also slow them down by imparting high barriers for transitioning between multiple deep traps. 
This work highlights the importance of the underlying cytoskeletal structure not just in delivering components to specific cellular regions but also in modulating the rate at which intracellular particles can encounter each other.

\section{Results}

\subsection{Search and encounter rates of run-and-tumble particles}

To explore the effect of particle motility on encounter kinetics, we first consider a simple minimal model for actively transported particles. Specifically, we assume that particles engage in `run-and-tumble' behavior \cite{rupprecht2016optimal,mori2020universal}, wherein they move in a straight line with velocity $v$ for an exponentially distributed run time with average $\tau_\text{run}$. The corresponding average run length is $\lambda = v \tau_\text{run}$. While some past studies of intracellular transport have described vesicular motion using a L\'evy flight model, with long-tailed polynomially distributed run-times \cite{bartumeus2002optimizing}, many others assume exponentially distributed runs for the sake of simplicity \cite{benichou2011intermittent,rupprecht2016optimal,loverdo2008enhanced,pangarkar2005dynamics}, as we do here. 
At the end of each run, the particles instantaneously pick a new direction uniformly at random and proceed to run in that direction. Pairs of such particles are confined in a two-dimensional disc of radius $R$, with a scattering boundary such that any particle which hits the boundary randomly selects a new heading within the disc.

This minimalist model interpolates between effectively diffusive and primarily ballistic motion, depending on whether the run length $\lambda$ is much shorter or much longer than the other length scales of interest. The effective diffusivity of a run-and-tumble particle can be defined as $D_\text{eff} = v \lambda / 2$, based on its mean squared displacement after many tumbles in the absence of confinement \cite{berg1983random}. Particles that engage in longer runs have a higher $D_\text{eff}$ and will spread through a large domain more rapidly.
However, this does not necessarily endow the particle with the ability to quickly find small targets \cite{rupprecht2016optimal,kahana2008active}.

\begin{figure*}[hbt!]
  \includegraphics[width=\textwidth]{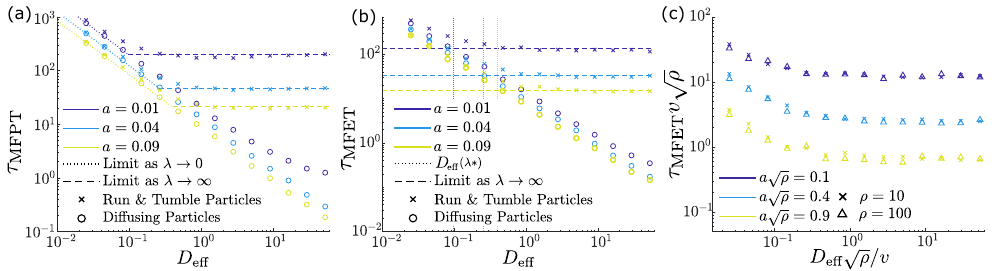}
\caption{
		Search and encounter of run-and-tumble versus diffusive particles. Length and timescales are nondimensionalized such that domain size $R=1$ and velocity $v=1$. (a) Mean first passage time to a centrally located fixed target is plotted versus the effective diffusivity $D_\text{eff}$ of run-and-tumble particles (crosses) and diffusive particles (circles), for three different particle diameters ($a$). The run length $\lambda$ of run-and-tumble particles is varied to yield different values of $D_\text{eff}$. Dotted line gives exact analytical solution for diffusive encounter with a central target (Eq.~\ref{eq:MFPT}). Dashed lines give the analytic $\lambda \rightarrow \infty$ limit (Eq.~\ref{eq:MFETlong}). (b) Analogous plots of encounter times between two identical particles engaging in run-and-tumble (crosses) or diffusive (circles) motion. Dashed lines give analytic $\lambda \rightarrow \infty$ limit. Dotted vertical lines correspond to the approximate critical run-length where processive motion slows down encounter (Eq.~\ref{lambda_star}). (c) Mean time to encounter the first of many particles present at density $\rho$. Data is plotted in dimensionless units, showing collapse for different particle densities (crosses: $\rho=10$, triangles: $\rho=100$).  See Supplemental Material for simulation details. }
	\label{fig:runtumble}
\end{figure*}

We note that analysis of single particle tracking data for intracellular motility often involves extracting the effective scaling and prefactor of the mean squared displacement~\cite{verkman2002solute, koslover2016disentangling,lin2016active}, so that $D_\text{eff}$ provides a minimal description of particle motility regardless of the underlying mechanism of motion. We therefore rely on $D_\text{eff}$ as the `control parameter' dictating how rapidly the particles are moving. 

In Fig.~\ref{fig:runtumble}a, we plot the average time to find a small fixed target of radius $a$ in the center of the domain, as a function of $D_\text{eff}$. 
As expected, active particles with very short run lengths have the same mean first passage time (MFPT) to the target as purely diffusive particles with the corresponding effective diffusivity. The diffusive target search time scales inversely with $D_\text{eff}$, as expected.
The simulated MFPT matches to the analytic solution \cite{le_vot2022first} for diffusive particles starting uniformly scattered in the domain, given by:
\begin{equation}
\mathcal{T}_D=\frac{R^4}{8D}\cdot\frac{4\ln(R/a)+4(a/R)^2-3-(a/R)^4}{R^2-a^2}.
\label{eq:MFPT}
\end{equation}

When the run length $\lambda$ of active particles increases, the MFPT to the target reaches a plateau. 
This effect can be explained by considering what happens as $\lambda\to\infty$. Once the average run length of the particles begins to approach the size of the domain, most particles encounter the domain boundary and pick a new direction before tumbling. Thus, the boundary effectively caps the run length for the run-and-tumble particles, so that more persistent motion no longer affects the search time.  Dimensional analysis indicates the limiting search time can be written as $\tau_\text{MFPT} = (R/v) f(a/R)$ (for some function $f$). More explicitly, we can calculate this search time 
by breaking up the particle trajectory into independent segments that start at the domain boundary and continue until either the target or the boundary is hit. This process can be treated as a form of `exploratory dynamics' with some splitting probability for either successfully hitting the target or resetting back to the boundary. As highlighted in recent work~\cite{kondev2025biological,koslover2025many}, the number of attempts prior to successfully hitting the target is given by the ratio of unsuccessful to successful runs: $n_\text{run} = (1-p_\text{hit})/p_\text{hit}$, where $p_\text{hit}$ is the probability of each independent run hitting the target. For the limit considered here, this probability can be computed as  $p_\text{hit} = (2/\pi) \sin^{-1}(a/R)$. 
Each of the unsuccessful runs has average length:
\begin{equation}
\begin{split}
	\left<\ell\right> = \frac{2(R-a)}{\pi/2 - \sin^{-1}(a/R)},	
\end{split}
\end{equation}
and the final successful run should have a length of approximately $R$. The average time to hit the target in the $\lambda\rightarrow \infty$ limit can then be approximated by
\begin{equation}
\begin{split}
	\tau_\text{MFPT} \rightarrow \frac{1-p_\text{hit}}{p_\text{hit}}\cdot\frac{\langle l\rangle}{v}+\frac{R}{v}.
\end{split}
\label{eq:MFETlong}
\end{equation}
As $\lambda$ increases, the search time asymptotes to this long-run limit, shown as dashed lines in Fig.~\ref{fig:runtumble}a. The transition where long run lengths become inefficient for finding the target, as compared to diffusive motion with the same dispersion rate $D_\text{eff}$ occurs when the two analytic approximations meet, at
\begin{equation}
\label{lambda_star}
{\lambda^*}\approx\frac{a(4\ln(R/a)-3)}{8}.
\end{equation}

\begin{figure}[hbt!]
	\centering
	\includegraphics[width=8.3cm]{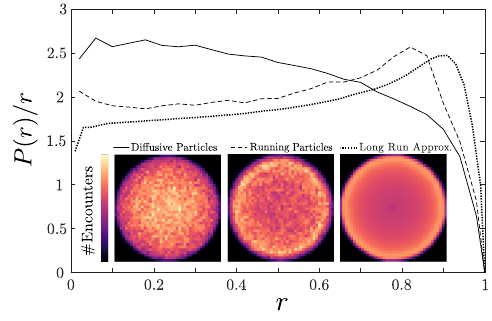}
	\caption{
		Distribution of encounter locations $P(r)$ for diffusing (solid)
		and running (dashed, $\lambda=1$) particles with diameter $a=0.02$. All length units are non-dimensionalized by domain radius $R=1$. The dotted line shows an analytic approximation for the encounter locations of running particles in the long-run limit. Inset shows spatial distributions of encounter locations for left: simulated diffusive particles, center: simulated run-and-tumble particles, right: analytic approximation for run-and-tumble particles. Simulation details provided in Supplemental Material.}
	\label{fig:encounterloc}
\end{figure}

Notably, this transition occurs at run-lengths on the order of the target size (much smaller than the domain size). For example, to encounter a $0.4\mu$m organelle within a mammalian cell or radius $R\approx 10\mu$m, the critical run length is approximately $\lambda^* \approx 0.5\mu$m.
This implies that in order to rapidly search for targets inside the the cell, such organelles gain no benefit from persistent runs that are longer than a micron or so. This is within the range of typical vesicular organelle run lengths of 0.2-10$\mu$m~\cite{shubeita2008consequences,ahmed2012mechanical,guimaraes2015peroxisomes,lin2016active}. We note that in past work, asymptotic expansion of the long-run-length limit demonstrated that the MFPT actually increases towards the asymptote with increasing $\lambda$, giving rise to an optimal run length~\cite{rupprecht2016optimal}. However, this increase is very shallow, so that the target search time at $\lambda^*$ is a reasonable approximation of the optimal MFPT.

In addition to the search time for a fixed central target, we consider the mean first encounter time (MFET) between two identically moving particles of diameter $a$ 
within the disc-shaped domain. As shown in Fig.~\ref{fig:runtumble}b, the MFET exhibits a similar behavior with respect to the effective diffusivity as does the search time for a stationary target. Small run-lengths yield the same encounter times as for diffusive particles, while long run lengths result in a limiting asymptote for the MFET. The approximate encounter time in the long-run limit is derived in the Supplemental Material.

We note that the MFET is approximately $20\%$ smaller than the search time for a fixed target, for both short and long run-lengths. Although the diffusivity of the mobile particles relative to each other is a factor of two higher when both of them are moving, the speed-up in the encounter is less substantial. This may be due to the placement of the fixed target in the most accessible region of the domain (the center), while the mobile particles spend time near the domain boundary where access to the other particle is partially blocked.

 The influence of domain boundaries on encounter times between moving particles has previously been explored in the context of particles exploring network structures~\cite{scott2023endoplasmic,weng2017hunting}. Two diffusing particles are more likely to find each other in the the center rather than near the boundary. For a one-dimensional domain, this effect can be demonstrated analytically (see Supplemental Material), and simulations indicate that an analogous preference for encounters in the bulk applies to diffusive particles in two dimensions as well (Fig.~\ref{fig:encounterloc}). Interestingly, run-and-tumble particles with long run-lengths show a very different distribution of encounter locations, with encounters occurring preferentially near the periphery of the domain. This effect can be approximated analytically in the limit of infinitely long runs (see Supplemental Material, dotted curve in Fig.~2).  The preference for peripheral encounters can be conceptually rationalized by breaking up the search process into a sequence of `attempts' in which one particle starts at a boundary and the other starts distributed throughout the domain. A successful attempt is more likely to terminate in encounter near a boundary, since one particle started there. An unsuccessful attempt ends when one of the particles hits the boundary, resetting the system to try again. This provides yet another manifestation of exploratory dynamics~\cite{kondev2025biological}, where the resetting process (hitting a boundary) modulates the steady-state distribution of the system~\cite{koslover2025many}.

Many cellular particles need to encounter one of a population of interaction partners, rather than a specific target. Examples include autophagosomes that must find a lysosome~\cite{cason2022spatiotemporal} or APPL1-bearing endosomes that must meet an EEA1-marked endosome~\cite{york2023deterministic} to proceed towards maturation. We therefore calculate the mean first encounter time for a fixed particle to find the first of many identical moving particles, present at a spatial density $\rho$. This density rescales the effective domain size that must be searched for the encounter. We can therefore plot the non-dimensionalized MFET ($\tau v\sqrt{\rho}$)
versus the non-dimensionalized effective diffusivity ($D_\text{eff} \sqrt{\rho}/v$) and target size ($a\sqrt{\rho}$). With this rescaling, different densities collapse together (Fig. 1c) to give an approximate universal behavior that describes how rapidly run-and-tumble particles find interaction partners among a population.
As with single-target encounters, increasing the run-length speeds up the encounter for short runs, but has little effect once the run-length is sufficiently long compared to the target.

\begin{figure*}
\centering
\includegraphics[width=\textwidth]{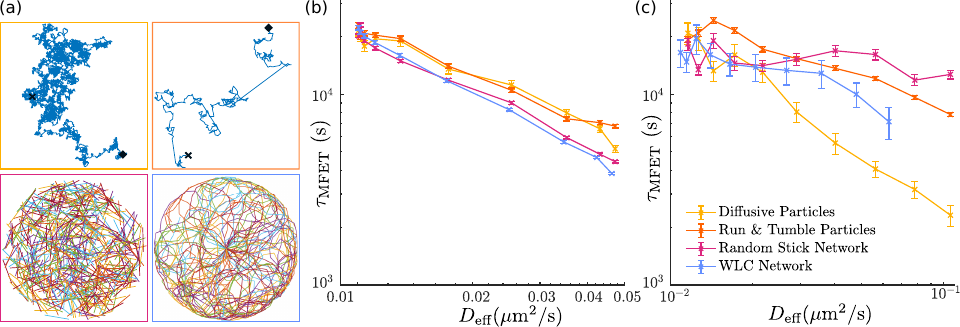}
\caption{
	Encounter times for unbiased particles exhibiting different modes of motion. (a) Top: example trajectories are shown for diffusive particles (yellow) and run-and-tumble particles (orange); bottom: example realizations of explicit, random-stick (pink) and worm-like aster (blue) networks. 
	(b) Mean first encounter time for pairs of particles corresponding to the motility modes illustrated in (a). Particles engaging in run-and-tumble motion or motion on explicit networks have short run lengths, with $\lambda = 0.1\mu$m. $D_\text{eff}$ is varied by changing $k_\text{on}$ in the range $10^{-1}-10^3 \text{s}^{-1}\mu\text{m}^{-1}$. (c) Analogous plots of MFET for particles with long run lengths, $\lambda = 4\mu$m. Error bars show standard error of the mean. The radius of the domain is $R=15\mu$m, the particle diameter is $a=0.2\mu$m, the random stick networks have filaments of length $5\mu$m and the worm-like aster has filaments of length $60\mu$m. Network density is $4.25 \mu$m$^{-1}$ in both cases, and the specific network structures used are illustrated in (a). Simulation details are provided in Supplemental Material.}		
\label{fig:fig3}
\end{figure*}

\subsection*{Encounters of Particles on Explicit Networks}
In cells, motor-driven organelles move along networks of cytoskeletal filaments that function as highways. The stability and organization of these networks can vary widely depending on the cell type~\cite{mogre2020getting,burute2019cellular}. Microtubules, which serve as the primary substrates for long-distance transport in animal cells, tend to emanate from one or more organizing centers~\cite{sallee2021microtubule} (MTOCs), and their effective intracellular flexbility allows for frequent intersections~\cite{balint2013correlative}. Actin filaments can make branched networks or random mesh structures~\cite{skau2015specification}, with some cargo switching between interpenetrating actin and microtubule networks~\cite{oberhofer2020molecular}. In past work, the density, polarity, and spatial organization of the network, together with rates of switching between distinct motility modes, were shown to play an important role in controlling the time required for particles to travel from the bulk to the periphery~\cite{ando2015cytoskeletal}, as well as the steady-state particle distribution~\cite{oberhofer2020molecular}.
Here, we consider explicit stable  network architectures as a source of quenched disorder underlying active particle motion, examining how encounter times between pairs of moving particles are affected by the network structure.

\subsubsection*{Unbiased Particles}
We simulate the motion of particles running on two different types of explicit networks: a `random stick' network (also known as a Mikado network~\cite{broedersz2014modeling}),
composed of short straight segments randomly scattered through the domain, and a worm-like-aster network consisting of semiflexible worm-like chains (WLCs) that emanate from a central point towards the cell periphery, bending and entangling when they reach the outer edge. 
The former can be taken to represent an actin mesh; the latter corresponds to microtubule structures observed in mammalian cells with a perinuclear MTOC. Both networks have a spatial density of $\rho \approx 4\mu\text{m}^{-1}$ (filament length per domain area), consistent with microtubule densities observed in fish melanophores \cite{burakov2021persistent}. 

Bound particles move along individual filaments with velocity $v$, and unbind from the filament with constant rate $k_\text{off}$. 
While unbound, they diffuse freely with diffusivity $D$. They are able to rebind to nearby filaments at rate $k_\text{on} \ell_a$, where $\ell_a$ is the length of the filament found within contact distance $a/2$ from the particle center. 

 The effective spatially averaged on-rate for particles on networks can be defined as $\left<k_\text{on}\right> = k_\text{on} \rho \pi a^2/4$.  
 The effective long-time diffusivity on a dense network of semiflexible chains can be approximated as (see Supplemental Material for details): 
 \begin{equation}
 \label{eq:network_deff}
 D_\text{eff}=\frac{Dk_\text{off}}{k_\text{off}+\left<k_\text{on}\right>}+\frac{v^2l_p\left<k_\text{on}\right>}{2(k_\text{off}l_p+v)(k_\text{off}+\left<k_\text{on}\right>)},
 \end{equation}
 where $\ell_p$ is the filament persistence length and the second term accounts for the limited persistence of particle directionality arising from bending of the tracks. Networks of random sticks lie in the limit $\ell_p \rightarrow \infty$, and the above approximation assumes that the length of each filament is long enough to not limit the run length (longer than $vk_\text{off}$).

We compute the mean first encounter times (MFET) for pairs of particles on explicit networks, with varying on-rates $k_\text{on}$. In Fig.~\ref{fig:fig3}, these encounter times are plotted against the effective diffusivity $D_\text{eff}$ and are compared to matched simulations of run-and-tumble or purely diffusive particles. The run-and-tumble particles are assumed to initiate runs with the rate $\left<k_\text{on}\right>$, run with velocity $v$, halt runs with the rate $k_\text{off}$, and diffuse with diffusivity $D$ until a new run is initiated. The purely diffusive particles are given a diffusivity of $D_\text{eff}$ for comparison.

We first consider the regime in which the run length $\lambda = 0.10\mu$m is small relative to the critical value $\lambda^* = 0.36\mu$m. As shown in Fig. \ref{fig:fig3}b,  such particles have similar MFETs regardless of their mode of motion or the network structure. The same similarity of short-length run-and-tumble motion to diffusive motion was also seen in Fig. \ref{fig:runtumble}. The presence of an explicit network reduces the encounter times slightly, an effect that can be attributed to particles being partially confined to the regions along individual filaments, requiring them to explore a total area smaller than the full domain before finding each other.
This effect increases when particles spend more time on the filaments (higher $k_\text{on}$, corresponding to higher $D_\text{eff}$).

In the regime of longer run lengths ($\lambda=4\mu$m), we observe that diffusive particles outperform both run \& tumble particles and particles on explicit networks (Fig. \ref{fig:fig3}c), particularly for high values of $D_\text{eff}$ (high $k_\text{on}$) where the active particles spend most of their time running.
This result is in keeping with the tendency of running particles to cover distances rapidly (high $D_\text{eff}$) yet overshoot a small target, as seen in Fig. \ref{fig:runtumble}. Once the run length becomes a substantial fraction of the domain size, the higher value of $D_\text{eff}$  does not actually increase the ability of the particles to search through the domain.

For particles that spend most of their time engaging in long runs (blue curve, Fig.~\ref{fig:fig3}c), encounter times are slighly lower on a network composed of worm-like chains than on one composed of straight sticks. This effect can be attributed to the bending of the chains allowing running particles to sample a wider region of space as compared to perfectly straight runs. In other words, the volume of the `Wiener sausage' (the area of space covered by a finite-sized particle undergoing a random walk)~\cite{berezhkovskii1989wiener} is expected to be higher for a particle moving along a wiggling chain rather than a straight one.                                                                                                                                                                                                                                                                                                          

Notably, encounter times of particles moving on predefined random-stick networks are similar to those engaged in spatially uniform run-and-tumble motion (Fig. \ref{fig:fig3}, right). For the physiologically relevant network density used here, particles encounter more than one filament during each diffusive phase of motion. The direction of each run is thus roughly uncorrelated to the previous one and the network can be well approximated by an isotropic continuum where particles can move in any direction after each tumble (or equivalently after each unbinding and rebinding cycle).
This suggests that unbiased organelle transport can be simplified to a run \& tumble model with few parameters: an effective diffusivity, an effective run length, and an encounter radius.

\begin{figure*}[h!]
\centering
\vspace{-0.5in}
\includegraphics[width=\textwidth]{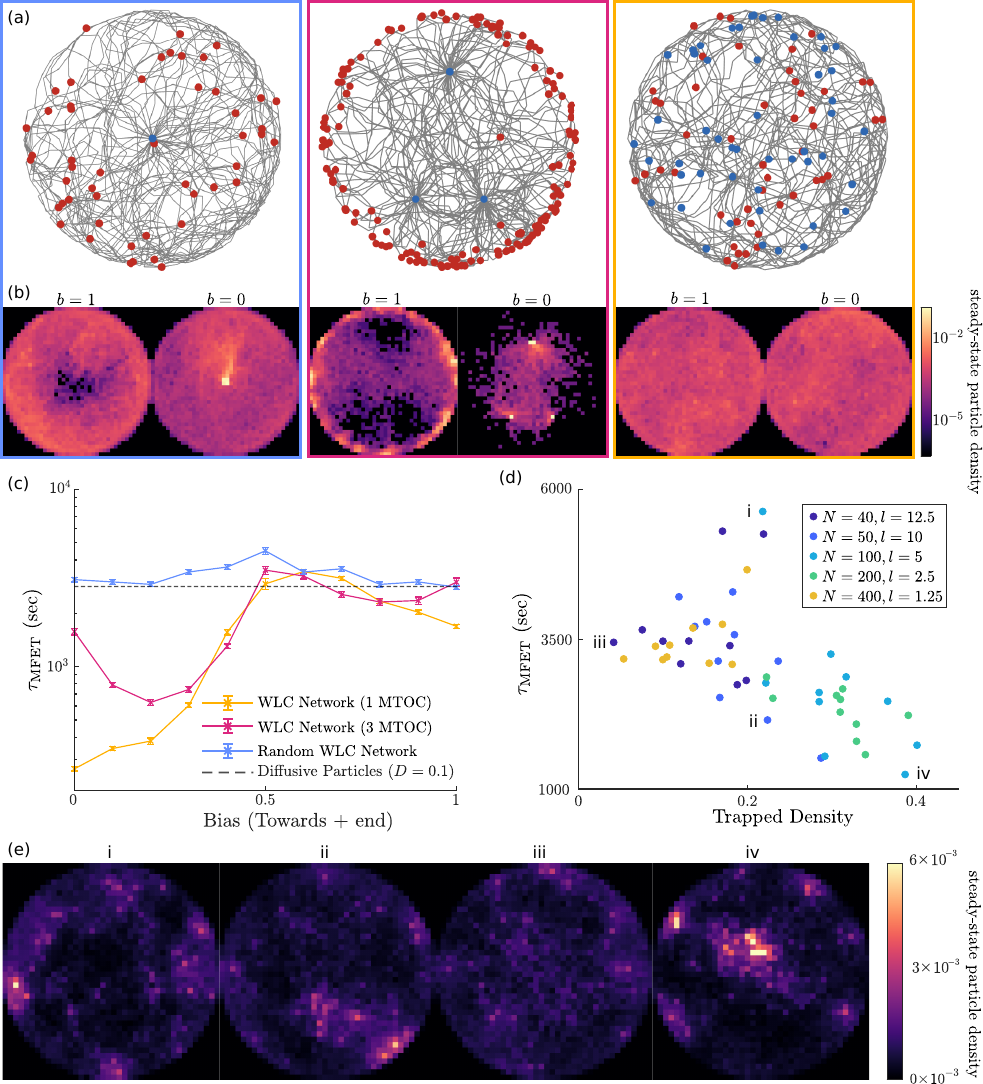}
\caption{
Encounter of biased particles on fixed network structures. (a) Example network structures for: (left) a WLC network with one MTOC and filament length 60 $\mu$m; (center) a WLC network with three MTOC and filaments extending to the domain boundary; (right) a randomly scattered WLC network with filament length 60 $\mu$m.  Red dots mark the plus end and blue dots mark the minus end for each filament. (b) Steady-state particle spatial distributions on the three networks shown in (a), for particles biased towards the plus end ($b=1$) and those biased towards the minus end ($b=0$). (c) Mean first encounter time plotted against particle bias for particles running on the three specific networks shown in (a). The horizontal dashed line shows the mean first encounter time for diffusive particles with the same effective diffusivity. (d) Mean first encounter time for particles with $b=1$ on individual instances of random WLC networks with different numbers and lengths of filaments, but fixed filament density. MFET is plotted versus the trapped fraction, defined as the fraction of particles that fall within dense clusters in the steady-state distribution. (e) Steady-state particle distributions are shown for four example networks in (d), indicating how positioning of traps can alter the encounter time. Parameters used throughout are: $a=0.2\mu\text{m}, k_\text{on} = 10\mu\text{m}^{-1}\text{s}^{-1}, k_\text{off} = 2\text{s}^{-1}, R=15\mu\text{m}$ and filament length density $4.25\mu\text{m}^{-1}$.
}
\label{fig:fig4}
\end{figure*}

\subsubsection*{Biased Particles}
Although many organelles exhibit bidirectional motion~\cite{hancock2014bidirectional}, there are also many cases where particle movements are biased, with a preference for moving towards either the plus or the minus end of each cytoskeletal filament. Such bias, which can be encoded by the recruitment or activation of directionally specific molecular motors, enables particles to be more efficiently delivered to specific cellular regions via spatially organized cytoskeletal structures~\cite{agrawal2022morphology}. For example, axonal microtubules are generally oriented with their plus-ends pointing outward from the cell body~\cite{kapitein2011way}, enabling the anterograde transport of presynaptic vesicles towards the distal tip and the retrograde transport of autophagosomes towards the cell body~\cite{guedes2019axonal}. In other cell types, multiple microtubule organizing centers~\cite{sallee2021microtubule} or randomly-oriented unpolarized actin networks~\cite{skau2015specification} may result in even biased particles moving in apparently random spatial directions. Here, we consider how quickly biased particles can encounter each other when moving along explicitly constructed spatial networks of filaments (Fig.~\ref{fig:fig4}a). 

We say that a particle has bias $b$ if, upon binding to a filament, it has a probability $b$ of running towards the plus end. Thus, unbiased particles have $b=0.5$, ones that always run towards the plus end have $b=1$, and ones that always run towards the minus end have $b=0$.

A key behavior observed for biased particles is their tendency to become `trapped' in localized regions where multiple microtubules converge with an inward orientation. Such traps can be seen as peaks in the steady-state distribution of non-interacting particles (Fig. \ref{fig:fig4}b,e). Modest trapping 
arises even in unstructured networks where filaments are scattered at random over the domain (the ``random WLC" networks in Fig. \ref{fig:fig4}a, right). The tendency towards trapping on random networks was previously shown to slow down the mean first passage time of biased particles to cross a rectangular region~\cite{mlynarczyk2019first} or move from a centralized nucleus to the outer periphery of the cell, particularly when the traps were localized near the nucleus~\cite{ando2015cytoskeletal}. Here, we find that the presence of traps can actually speed up the encounter time between particles (Fig. \ref{fig:fig4}c,d).

One limiting regime is a fully polarized network where all filaments originate with their minus ends attached to a single central point, representing a MTOC. In this case, the MFET for particles biased to run towards the minus end ($b=0$) is very short, since they quickly encounter each other at the MTOC (Fig. \ref{fig:fig4}c).
A bias towards the peripherally-oriented plus-ends of the network also slightly speeds up the encounter time as compared to unbiased particles. Particles moving towards the plus-end do not converge to a single point-like trap, but rather move rapidly towards a narrow region at the periphery of the cell. Within that boundary layer, they undergo random walks in the angular coordinate as they unbind, diffuse, and hop between the filaments. These results are analogous to those observed in spatially heterogeneous random velocity models, where the imposition of inward or outward bias in the bulk of the domain was also shown to speed up encounters~\cite{hafner2018spatial}.
The ability to find a partner is  enhanced because the particles need only perform an effectively one-dimensional search in a narrow boundary layer, rather than a two-dimensional search in the whole domain. 
Such enhancement in search times through dimensional reduction has been noted in several contexts~\cite{koslover2024searching}, including diffusive target search by particles in yeast ER networks,  facilitated diffusion of proteins that bind and slide along a DNA chain, and reaction rates of particles that interact weakly with membranes.

The effect of bias on encounter times becomes more complicated when the network configuration includes multiple traps. In Fig. \ref{fig:fig4}c (pink curve) we show the MFET for a network of filaments growing from three distinct MTOCs placed halfway between the center of the domain and the boundary. In this case, a small bias towards the negative tip ($b<0.5$) enhances encounter rates by funneling particles to a smaller sub-regions of the domain. However, excess bias ($b\rightarrow 0$) slows down the encounter. 
In networks with multiple deep traps, a biased particle may become ``stuck" in a different trap from its partner. In order to find each other, one of the particles must escape its trap, which greatly increases the encounter time for strongly biased particles.

In addition, when filaments are grown from distinct organizing centers, their plus ends tend to be non-uniformly distributed along the periphery, forming clusters separated by gaps with few filament tips (Fig. \ref{fig:fig4}a, center). As a result, particles that are strongly biased towards the plus ends ($b \rightarrow 1$), while benefiting from the dimensional reduction effect described previously, can also become partially trapped within distinct peripheral subregions, increasing their encounter times.

Networks with randomly placed filaments (Fig.~\ref{fig:fig4}a, right) also exhibit multiple traps for biased particles. However, these traps are much shallower than the ones formed by filaments emanating from specific organizing centers (Fig.~\ref{fig:fig4}b). Biased particles running on random networks are thus able to transition between traps more easily. The search process for biased particles can be thought of as a series of discrete hops between a finite number of trapped states. Deep traps formed by converging filaments give rise to exponentially slow transition times between such states, as previously noted for one-dimensional systems ~\cite{sarpangala2024tunable}. For the random network, more states are available and the transition times between them are much lower. Thus, the encounter times on these networks are less sensitive to bias in the particle runs.

To further dissect the role of trapping in particle encounter times, we compare MFETs on individual network structures (composed of randomly distributed straight filaments). The total density of the networks is fixed, while the length and number of filaments is allowed to vary. For each network, we compute the `trapped density' by first allowing the simulation to run until the distribution of particles reaches a steady state (Fig.~\ref{fig:fig4}d). We then perform a clustering analysis using the DBSCAN algorithm~\cite{Ester1996DBSCAN} on sampled particle positions
to break up the populations into dense clusters versus broadly scattered particles (details in Supplemental Material). The trapped density is defined as the fraction of all particles found within the dense clusters.
As seen in Fig.~\ref{fig:fig4}d, networks where particles are more likely to be trapped generally allow for faster encounter times. Trapping on a random network confines the particles to smaller regions of space without imposing prohibitively large transition times, allowing the particles to more rapidly find each other. Interestingly, networks with intermediate trapped densities show the greatest variability in mean encounter times, with the MFET varying more than two-fold depending on the specific network structure. For a given overall trapped density, structures which split the particles up into multiple distant traps (Fig.~\ref{fig:fig4}e.i) tend to give the highest MFET whereas those where traps cluster together to form a single localized basin (Fig.~\ref{fig:fig4}e.ii) yield the fastest encounter times.

We note that particles are more likely to be found in traps when the network is composed of filaments with intermediate lengths. When filaments are very long, there are few tips and regions of convergence are less likely to form. In the case of many very short filaments, the network begins to look homogeneous and isotropic, and large-scale traps are less likely to appear. Intermediate filament lengths allow the most significant traps to form, simultaneously providing enough tips to form clusters and a sufficiently large basin of convergence into those clusters to result in substantial trapping. As a result, the fastest encounter times are observed for random networks formed with intermediate filament lengths.

\section{Conclusions}

Many cellular processes rely on `exploratory dynamics'~\cite{kondev2025biological}, wherein components move through space blindly until they encounter a target or binding partner that allows them to carry out their function. Often, both reaction partners are mobile, as in the case of interactions between intermittently motor-driven vesicular organelles. The timing of such encounters serves as a regulatory mode for controlling processes such as organelle maturation or signal persistence in space and time~\cite{koslover2024searching}. Intracellular particles move through a complex landscape of filaments that support locally directed transport but can hinder mixing via the formation of local traps. Understanding interaction dynamics at the organelle and cellular scales requires unraveling how the interplay between particle kinetics and filament organization governs encounter times.

In this work we explore first encounter times between pairs of particles that engage in intermittent processive and diffusive motion, both with and without an underlying fixed filament network.
Increasing the run-length of such particles enhances their ability to search through unbounded space but leads to an asymptotic non-zero value for the encounter time. Thus, diffusive particles with an equivalent spreading rate are able to find each other more rapidly than particles with long run-lengths. In an unstructured domain, encounters between diffusive particles are more likely to occur in the bulk of the domain, while those with long run lengths are more likely to find each other near the periphery.

We next consider the effect of an explicit underlying network structure, which serves as a source of quenched disorder for particle motion. Unbiased particles moving on dense networks behave similarly to run-and-tumble particles in an empty domain. For biased particles, by contrast, explicit network architectures give rise to traps, which have previously been shown to slow down overall transport rates~\cite{ando2015cytoskeletal,mlynarczyk2019first}. We demonstrate that the existence of traps can actually decrease encounter times between pairs of moving particles, by funneling them jointly into trapped regions, reducing the dimensionality of the space that the particles need to explore.

The models presented here constitute a highly simplified description of intracellular transport. In particular, we assume exponentially distributed transition times, while some studies indicate that vesicular organelles may undergo L\'evy flights with long-tailed run length distributions~\cite{chen2015memoryless,fedotov2018memory}. The underlying filament network structure is taken to be a fixed architecture, neglecting the turnover of microtubules and actin filaments in living cells~\cite{schulze1986microtubule,fritzsche2013analysis}. Furthermore, the motion of intracellular particles can be influenced by other spatially distributed features beyond the network architecture, including cytoplasmic obstacles, microtubule-associated proteins, and post-translational modifications to cytoskeletal filaments~\cite{mogre2020getting}. Nevertheless, the model framework described here provides a simple basis for  
connecting network architecture and transport parameters with the kinetics of interparticle encounters.

These results provide a framework for estimating how quickly cellular particles can find each other starting from observable parameters such as their effective diffusivity, typical run time, and the underlying network structure. It should be noted that limitations in spatiotemporal resolution often imply that only the effective time- and ensemble-averaged motion of the particles can be easily extracted from experimental data. Our numerical model makes it possible to estimate encounter rates starting from this effective behavior. In addition, the model highlights the importance of both network structure and particle bias, implicating these features as potentially important for regulating the timing of cellular processes that rely on organelle encounters.

\clearpage

\bibliography{encountertimes.bib}

\begin{thebibliography}{76}%
\makeatletter
\providecommand \@ifxundefined [1]{%
 \@ifx{#1\undefined}
}%
\providecommand \@ifnum [1]{%
 \ifnum #1\expandafter \@firstoftwo
 \else \expandafter \@secondoftwo
 \fi
}%
\providecommand \@ifx [1]{%
 \ifx #1\expandafter \@firstoftwo
 \else \expandafter \@secondoftwo
 \fi
}%
\providecommand \natexlab [1]{#1}%
\providecommand \enquote  [1]{``#1''}%
\providecommand \bibnamefont  [1]{#1}%
\providecommand \bibfnamefont [1]{#1}%
\providecommand \citenamefont [1]{#1}%
\providecommand \href@noop [0]{\@secondoftwo}%
\providecommand \href [0]{\begingroup \@sanitize@url \@href}%
\providecommand \@href[1]{\@@startlink{#1}\@@href}%
\providecommand \@@href[1]{\endgroup#1\@@endlink}%
\providecommand \@sanitize@url [0]{\catcode `\\12\catcode `\$12\catcode
  `\&12\catcode `\#12\catcode `\^12\catcode `\_12\catcode `\%12\relax}%
\providecommand \@@startlink[1]{}%
\providecommand \@@endlink[0]{}%
\providecommand \url  [0]{\begingroup\@sanitize@url \@url }%
\providecommand \@url [1]{\endgroup\@href {#1}{\urlprefix }}%
\providecommand \urlprefix  [0]{URL }%
\providecommand \Eprint [0]{\href }%
\providecommand \doibase [0]{https://doi.org/}%
\providecommand \selectlanguage [0]{\@gobble}%
\providecommand \bibinfo  [0]{\@secondoftwo}%
\providecommand \bibfield  [0]{\@secondoftwo}%
\providecommand \translation [1]{[#1]}%
\providecommand \BibitemOpen [0]{}%
\providecommand \bibitemStop [0]{}%
\providecommand \bibitemNoStop [0]{.\EOS\space}%
\providecommand \EOS [0]{\spacefactor3000\relax}%
\providecommand \BibitemShut  [1]{\csname bibitem#1\endcsname}%
\let\auto@bib@innerbib\@empty
\bibitem [{\citenamefont {Koslover}(2024)}]{koslover2024searching}%
  \BibitemOpen
  \bibfield  {author} {\bibinfo {author} {\bibfnamefont {E.~F.}\ \bibnamefont
  {Koslover}},\ }\bibfield  {title} {\bibinfo {title} {Searching through
  cellular landscapes},\ }in\ \href@noop {} {\emph {\bibinfo {booktitle}
  {Target Search Problems}}}\ (\bibinfo  {publisher} {Springer},\ \bibinfo
  {year} {2024})\ pp.\ \bibinfo {pages} {541--577}\BibitemShut {NoStop}%
\bibitem [{\citenamefont {Fourriere}\ \emph {et~al.}(2020)\citenamefont
  {Fourriere}, \citenamefont {Jimenez}, \citenamefont {Perez},\ and\
  \citenamefont {Boncompain}}]{fourriere2020role}%
  \BibitemOpen
  \bibfield  {author} {\bibinfo {author} {\bibfnamefont {L.}~\bibnamefont
  {Fourriere}}, \bibinfo {author} {\bibfnamefont {A.~J.}\ \bibnamefont
  {Jimenez}}, \bibinfo {author} {\bibfnamefont {F.}~\bibnamefont {Perez}},\
  and\ \bibinfo {author} {\bibfnamefont {G.}~\bibnamefont {Boncompain}},\
  }\bibfield  {title} {\bibinfo {title} {The role of microtubules in secretory
  protein transport},\ }\href@noop {} {\bibfield  {journal} {\bibinfo
  {journal} {J Cell Sci}\ }\textbf {\bibinfo {volume} {133}},\ \bibinfo {pages}
  {jcs237016} (\bibinfo {year} {2020})}\BibitemShut {NoStop}%
\bibitem [{\citenamefont {Sorkin}\ and\ \citenamefont
  {Von~Zastrow}(2009)}]{sorkin2009endocytosis}%
  \BibitemOpen
  \bibfield  {author} {\bibinfo {author} {\bibfnamefont {A.}~\bibnamefont
  {Sorkin}}\ and\ \bibinfo {author} {\bibfnamefont {M.}~\bibnamefont
  {Von~Zastrow}},\ }\bibfield  {title} {\bibinfo {title} {Endocytosis and
  signalling: intertwining molecular networks},\ }\href@noop {} {\bibfield
  {journal} {\bibinfo  {journal} {Nat Rev Mol Cell Bio}\ }\textbf {\bibinfo
  {volume} {10}},\ \bibinfo {pages} {609} (\bibinfo {year} {2009})}\BibitemShut
  {NoStop}%
\bibitem [{\citenamefont {Bakker}\ \emph {et~al.}(2017)\citenamefont {Bakker},
  \citenamefont {Spits}, \citenamefont {Neefjes},\ and\ \citenamefont
  {Berlin}}]{bakker2017egfr}%
  \BibitemOpen
  \bibfield  {author} {\bibinfo {author} {\bibfnamefont {J.}~\bibnamefont
  {Bakker}}, \bibinfo {author} {\bibfnamefont {M.}~\bibnamefont {Spits}},
  \bibinfo {author} {\bibfnamefont {J.}~\bibnamefont {Neefjes}},\ and\ \bibinfo
  {author} {\bibfnamefont {I.}~\bibnamefont {Berlin}},\ }\bibfield  {title}
  {\bibinfo {title} {The egfr odyssey--from activation to destruction in space
  and time},\ }\href@noop {} {\bibfield  {journal} {\bibinfo  {journal} {J Cell
  Sci}\ }\textbf {\bibinfo {volume} {130}},\ \bibinfo {pages} {4087} (\bibinfo
  {year} {2017})}\BibitemShut {NoStop}%
\bibitem [{\citenamefont {York}\ \emph {et~al.}(2023)\citenamefont {York},
  \citenamefont {Joshi}, \citenamefont {Wright}, \citenamefont {Kreplin},
  \citenamefont {Rodgers}, \citenamefont {Moorthi}, \citenamefont {Gandhi},
  \citenamefont {Patil}, \citenamefont {Mitchell}, \citenamefont
  {Iyer-Biswas},\ and\ \citenamefont {Arumugam}}]{york2023deterministic}%
  \BibitemOpen
  \bibfield  {author} {\bibinfo {author} {\bibfnamefont {H.~M.}\ \bibnamefont
  {York}}, \bibinfo {author} {\bibfnamefont {K.}~\bibnamefont {Joshi}},
  \bibinfo {author} {\bibfnamefont {C.~S.}\ \bibnamefont {Wright}}, \bibinfo
  {author} {\bibfnamefont {L.~Z.}\ \bibnamefont {Kreplin}}, \bibinfo {author}
  {\bibfnamefont {S.~J.}\ \bibnamefont {Rodgers}}, \bibinfo {author}
  {\bibfnamefont {U.~K.}\ \bibnamefont {Moorthi}}, \bibinfo {author}
  {\bibfnamefont {H.}~\bibnamefont {Gandhi}}, \bibinfo {author} {\bibfnamefont
  {A.}~\bibnamefont {Patil}}, \bibinfo {author} {\bibfnamefont {C.~A.}\
  \bibnamefont {Mitchell}}, \bibinfo {author} {\bibfnamefont {S.}~\bibnamefont
  {Iyer-Biswas}},\ and\ \bibinfo {author} {\bibfnamefont {S.}~\bibnamefont
  {Arumugam}},\ }\bibfield  {title} {\bibinfo {title} {Deterministic early
  endosomal maturations emerge from a stochastic trigger-and-convert
  mechanism},\ }\bibfield  {journal} {\bibinfo  {journal} {Nat Commun}\
  }\textbf {\bibinfo {volume} {14}},\ \href
  {https://doi.org/10.1038/s41467-023-40428-1} {10.1038/s41467-023-40428-1}
  (\bibinfo {year} {2023})\BibitemShut {NoStop}%
\bibitem [{\citenamefont {Cason}\ \emph {et~al.}(2022)\citenamefont {Cason},
  \citenamefont {Mogre}, \citenamefont {Holzbaur},\ and\ \citenamefont
  {Koslover}}]{cason2022spatiotemporal}%
  \BibitemOpen
  \bibfield  {author} {\bibinfo {author} {\bibfnamefont {S.~E.}\ \bibnamefont
  {Cason}}, \bibinfo {author} {\bibfnamefont {S.~S.}\ \bibnamefont {Mogre}},
  \bibinfo {author} {\bibfnamefont {E.~L.}\ \bibnamefont {Holzbaur}},\ and\
  \bibinfo {author} {\bibfnamefont {E.~F.}\ \bibnamefont {Koslover}},\
  }\bibfield  {title} {\bibinfo {title} {Spatiotemporal analysis of axonal
  autophagosome--lysosome dynamics reveals limited fusion events and slow
  maturation},\ }\href@noop {} {\bibfield  {journal} {\bibinfo  {journal} {Mol
  Biol Cell}\ }\textbf {\bibinfo {volume} {33}},\ \bibinfo {pages} {ar123}
  (\bibinfo {year} {2022})}\BibitemShut {NoStop}%
\bibitem [{\citenamefont {Halford}\ and\ \citenamefont
  {Marko}(2004)}]{halford2004site}%
  \BibitemOpen
  \bibfield  {author} {\bibinfo {author} {\bibfnamefont {S.~E.}\ \bibnamefont
  {Halford}}\ and\ \bibinfo {author} {\bibfnamefont {J.~F.}\ \bibnamefont
  {Marko}},\ }\bibfield  {title} {\bibinfo {title} {How do site-specific
  dna-binding proteins find their targets?},\ }\href@noop {} {\bibfield
  {journal} {\bibinfo  {journal} {Nucleic Acids Res}\ }\textbf {\bibinfo
  {volume} {32}},\ \bibinfo {pages} {3040} (\bibinfo {year}
  {2004})}\BibitemShut {NoStop}%
\bibitem [{\citenamefont {Herbst}\ and\ \citenamefont
  {Martin}(2017)}]{herbst2017regulated}%
  \BibitemOpen
  \bibfield  {author} {\bibinfo {author} {\bibfnamefont {W.~A.}\ \bibnamefont
  {Herbst}}\ and\ \bibinfo {author} {\bibfnamefont {K.~C.}\ \bibnamefont
  {Martin}},\ }\bibfield  {title} {\bibinfo {title} {Regulated transport of
  signaling proteins from synapse to nucleus},\ }\href@noop {} {\bibfield
  {journal} {\bibinfo  {journal} {Curr Opin Neurobiol}\ }\textbf {\bibinfo
  {volume} {45}},\ \bibinfo {pages} {78} (\bibinfo {year} {2017})}\BibitemShut
  {NoStop}%
\bibitem [{\citenamefont {Devaux}\ \emph {et~al.}(2010)\citenamefont {Devaux},
  \citenamefont {Lelandais}, \citenamefont {Garcia}, \citenamefont {Goussard},\
  and\ \citenamefont {Jacq}}]{devaux2010posttranscriptional}%
  \BibitemOpen
  \bibfield  {author} {\bibinfo {author} {\bibfnamefont {F.}~\bibnamefont
  {Devaux}}, \bibinfo {author} {\bibfnamefont {G.}~\bibnamefont {Lelandais}},
  \bibinfo {author} {\bibfnamefont {M.}~\bibnamefont {Garcia}}, \bibinfo
  {author} {\bibfnamefont {S.}~\bibnamefont {Goussard}},\ and\ \bibinfo
  {author} {\bibfnamefont {C.}~\bibnamefont {Jacq}},\ }\bibfield  {title}
  {\bibinfo {title} {Posttranscriptional control of mitochondrial biogenesis:
  spatio-temporal regulation of the protein import process},\ }\href@noop {}
  {\bibfield  {journal} {\bibinfo  {journal} {Febs Lett}\ }\textbf {\bibinfo
  {volume} {584}},\ \bibinfo {pages} {4273} (\bibinfo {year}
  {2010})}\BibitemShut {NoStop}%
\bibitem [{\citenamefont {Stehbens}\ and\ \citenamefont
  {Wittmann}(2012)}]{stehbens2012targeting}%
  \BibitemOpen
  \bibfield  {author} {\bibinfo {author} {\bibfnamefont {S.}~\bibnamefont
  {Stehbens}}\ and\ \bibinfo {author} {\bibfnamefont {T.}~\bibnamefont
  {Wittmann}},\ }\bibfield  {title} {\bibinfo {title} {Targeting and transport:
  how microtubules control focal adhesion dynamics},\ }\href@noop {} {\bibfield
   {journal} {\bibinfo  {journal} {J Cell Biol}\ }\textbf {\bibinfo {volume}
  {198}},\ \bibinfo {pages} {481} (\bibinfo {year} {2012})}\BibitemShut
  {NoStop}%
\bibitem [{\citenamefont {Kondev}\ \emph {et~al.}(2025)\citenamefont {Kondev},
  \citenamefont {Kirschner}, \citenamefont {Garcia}, \citenamefont {Salmon},\
  and\ \citenamefont {Phillips}}]{kondev2025biological}%
  \BibitemOpen
  \bibfield  {author} {\bibinfo {author} {\bibfnamefont {J.}~\bibnamefont
  {Kondev}}, \bibinfo {author} {\bibfnamefont {M.}~\bibnamefont {Kirschner}},
  \bibinfo {author} {\bibfnamefont {H.~G.}\ \bibnamefont {Garcia}}, \bibinfo
  {author} {\bibfnamefont {G.~L.}\ \bibnamefont {Salmon}},\ and\ \bibinfo
  {author} {\bibfnamefont {R.}~\bibnamefont {Phillips}},\ }\bibfield  {title}
  {\bibinfo {title} {Biological processes as exploratory dynamics},\
  }\href@noop {} {\bibfield  {journal} {\bibinfo  {journal} {Biophys J}\ }
  (\bibinfo {year} {2025})}\BibitemShut {NoStop}%
\bibitem [{\citenamefont {Chustecki}\ \emph {et~al.}(2021)\citenamefont
  {Chustecki}, \citenamefont {Gibbs}, \citenamefont {Bassel},\ and\
  \citenamefont {Johnston}}]{chustecki2021network}%
  \BibitemOpen
  \bibfield  {author} {\bibinfo {author} {\bibfnamefont {J.~M.}\ \bibnamefont
  {Chustecki}}, \bibinfo {author} {\bibfnamefont {D.~J.}\ \bibnamefont
  {Gibbs}}, \bibinfo {author} {\bibfnamefont {G.~W.}\ \bibnamefont {Bassel}},\
  and\ \bibinfo {author} {\bibfnamefont {I.~G.}\ \bibnamefont {Johnston}},\
  }\bibfield  {title} {\bibinfo {title} {Network analysis of arabidopsis
  mitochondrial dynamics reveals a resolved tradeoff between physical
  distribution and social connectivity},\ }\href@noop {} {\bibfield  {journal}
  {\bibinfo  {journal} {Cell Syst}\ }\textbf {\bibinfo {volume} {12}},\
  \bibinfo {pages} {419} (\bibinfo {year} {2021})}\BibitemShut {NoStop}%
\bibitem [{\citenamefont {Holt}\ \emph {et~al.}(2025)\citenamefont {Holt},
  \citenamefont {Zurita}, \citenamefont {Teryoshin}, \citenamefont {Lewis},\
  and\ \citenamefont {Koslover}}]{holt2025diffusive}%
  \BibitemOpen
  \bibfield  {author} {\bibinfo {author} {\bibfnamefont {K.~B.}\ \bibnamefont
  {Holt}}, \bibinfo {author} {\bibfnamefont {C.}~\bibnamefont {Zurita}},
  \bibinfo {author} {\bibfnamefont {L.}~\bibnamefont {Teryoshin}}, \bibinfo
  {author} {\bibfnamefont {S.~C.}\ \bibnamefont {Lewis}},\ and\ \bibinfo
  {author} {\bibfnamefont {E.~F.}\ \bibnamefont {Koslover}},\ }\bibfield
  {title} {\bibinfo {title} {Diffusive spreading across dynamic mitochondrial
  network architectures},\ }\href@noop {} {\bibfield  {journal} {\bibinfo
  {journal} {arXiv preprint arXiv:2506.05643}\ } (\bibinfo {year}
  {2025})}\BibitemShut {NoStop}%
\bibitem [{\citenamefont {Lin}\ \emph {et~al.}(2016)\citenamefont {Lin},
  \citenamefont {Schuster}, \citenamefont {Guimaraes}, \citenamefont {Ashwin},
  \citenamefont {Schrader}, \citenamefont {Metz}, \citenamefont {Hacker},
  \citenamefont {Gurr},\ and\ \citenamefont {Steinberg}}]{lin2016active}%
  \BibitemOpen
  \bibfield  {author} {\bibinfo {author} {\bibfnamefont {C.}~\bibnamefont
  {Lin}}, \bibinfo {author} {\bibfnamefont {M.}~\bibnamefont {Schuster}},
  \bibinfo {author} {\bibfnamefont {S.~C.}\ \bibnamefont {Guimaraes}}, \bibinfo
  {author} {\bibfnamefont {P.}~\bibnamefont {Ashwin}}, \bibinfo {author}
  {\bibfnamefont {M.}~\bibnamefont {Schrader}}, \bibinfo {author}
  {\bibfnamefont {J.}~\bibnamefont {Metz}}, \bibinfo {author} {\bibfnamefont
  {C.}~\bibnamefont {Hacker}}, \bibinfo {author} {\bibfnamefont {S.~J.}\
  \bibnamefont {Gurr}},\ and\ \bibinfo {author} {\bibfnamefont
  {G.}~\bibnamefont {Steinberg}},\ }\bibfield  {title} {\bibinfo {title}
  {Active diffusion and microtubule-based transport oppose myosin forces to
  position organelles in cells},\ }\href@noop {} {\bibfield  {journal}
  {\bibinfo  {journal} {Nat Commun}\ }\textbf {\bibinfo {volume} {7}},\
  \bibinfo {pages} {11814} (\bibinfo {year} {2016})}\BibitemShut {NoStop}%
\bibitem [{\citenamefont {Drechsler}\ \emph {et~al.}(2017)\citenamefont
  {Drechsler}, \citenamefont {Giavazzi}, \citenamefont {Cerbino},\ and\
  \citenamefont {Palacios}}]{drechsler2017active}%
  \BibitemOpen
  \bibfield  {author} {\bibinfo {author} {\bibfnamefont {M.}~\bibnamefont
  {Drechsler}}, \bibinfo {author} {\bibfnamefont {F.}~\bibnamefont {Giavazzi}},
  \bibinfo {author} {\bibfnamefont {R.}~\bibnamefont {Cerbino}},\ and\ \bibinfo
  {author} {\bibfnamefont {I.~M.}\ \bibnamefont {Palacios}},\ }\bibfield
  {title} {\bibinfo {title} {Active diffusion and advection in drosophila
  oocytes result from the interplay of actin and microtubules},\ }\href@noop {}
  {\bibfield  {journal} {\bibinfo  {journal} {Nat Commun}\ }\textbf {\bibinfo
  {volume} {8}},\ \bibinfo {pages} {1520} (\bibinfo {year} {2017})}\BibitemShut
  {NoStop}%
\bibitem [{\citenamefont {Carlini}\ \emph {et~al.}(2020)\citenamefont
  {Carlini}, \citenamefont {Brittingham}, \citenamefont {Holt},\ and\
  \citenamefont {Kapoor}}]{carlini2020microtubules}%
  \BibitemOpen
  \bibfield  {author} {\bibinfo {author} {\bibfnamefont {L.}~\bibnamefont
  {Carlini}}, \bibinfo {author} {\bibfnamefont {G.~P.}\ \bibnamefont
  {Brittingham}}, \bibinfo {author} {\bibfnamefont {L.~J.}\ \bibnamefont
  {Holt}},\ and\ \bibinfo {author} {\bibfnamefont {T.~M.}\ \bibnamefont
  {Kapoor}},\ }\bibfield  {title} {\bibinfo {title} {Microtubules enhance
  mesoscale effective diffusivity in the crowded metaphase cytoplasm},\
  }\href@noop {} {\bibfield  {journal} {\bibinfo  {journal} {Dev Cell}\
  }\textbf {\bibinfo {volume} {54}},\ \bibinfo {pages} {574} (\bibinfo {year}
  {2020})}\BibitemShut {NoStop}%
\bibitem [{\citenamefont {Magiera}\ \emph {et~al.}(2025)\citenamefont
  {Magiera}, \citenamefont {Kucharska}, \citenamefont {Kalwarczyk},
  \citenamefont {Haniewicz}, \citenamefont {Kwapiszewska},\ and\ \citenamefont
  {Ho{\l}yst}}]{magiera2025measurement}%
  \BibitemOpen
  \bibfield  {author} {\bibinfo {author} {\bibfnamefont {A.}~\bibnamefont
  {Magiera}}, \bibinfo {author} {\bibfnamefont {K.}~\bibnamefont {Kucharska}},
  \bibinfo {author} {\bibfnamefont {T.}~\bibnamefont {Kalwarczyk}}, \bibinfo
  {author} {\bibfnamefont {P.}~\bibnamefont {Haniewicz}}, \bibinfo {author}
  {\bibfnamefont {K.}~\bibnamefont {Kwapiszewska}},\ and\ \bibinfo {author}
  {\bibfnamefont {R.}~\bibnamefont {Ho{\l}yst}},\ }\bibfield  {title} {\bibinfo
  {title} {Measurement of large ribosomal subunit size in cytoplasm and nucleus
  of living human cells},\ }\href@noop {} {\bibfield  {journal} {\bibinfo
  {journal} {Nanoscale Horiz}\ }\textbf {\bibinfo {volume} {10}},\ \bibinfo
  {pages} {388} (\bibinfo {year} {2025})}\BibitemShut {NoStop}%
\bibitem [{\citenamefont {Koslover}\ \emph {et~al.}(2016)\citenamefont
  {Koslover}, \citenamefont {Chan},\ and\ \citenamefont
  {Theriot}}]{koslover2016disentangling}%
  \BibitemOpen
  \bibfield  {author} {\bibinfo {author} {\bibfnamefont {E.~F.}\ \bibnamefont
  {Koslover}}, \bibinfo {author} {\bibfnamefont {C.~K.}\ \bibnamefont {Chan}},\
  and\ \bibinfo {author} {\bibfnamefont {J.~A.}\ \bibnamefont {Theriot}},\
  }\bibfield  {title} {\bibinfo {title} {Disentangling random motion and flow
  in a complex medium},\ }\href@noop {} {\bibfield  {journal} {\bibinfo
  {journal} {Biophys J}\ }\textbf {\bibinfo {volume} {110}},\ \bibinfo {pages}
  {700} (\bibinfo {year} {2016})}\BibitemShut {NoStop}%
\bibitem [{\citenamefont {Mogre}\ \emph {et~al.}(2020)\citenamefont {Mogre},
  \citenamefont {Brown},\ and\ \citenamefont {Koslover}}]{mogre2020getting}%
  \BibitemOpen
  \bibfield  {author} {\bibinfo {author} {\bibfnamefont {S.~S.}\ \bibnamefont
  {Mogre}}, \bibinfo {author} {\bibfnamefont {A.~I.}\ \bibnamefont {Brown}},\
  and\ \bibinfo {author} {\bibfnamefont {E.~F.}\ \bibnamefont {Koslover}},\
  }\bibfield  {title} {\bibinfo {title} {Getting around the cell: physical
  transport in the intracellular world},\ }\href@noop {} {\bibfield  {journal}
  {\bibinfo  {journal} {Phys Biol}\ }\textbf {\bibinfo {volume} {17}},\
  \bibinfo {pages} {061003} (\bibinfo {year} {2020})}\BibitemShut {NoStop}%
\bibitem [{\citenamefont {Nielsen}\ \emph {et~al.}(1999)\citenamefont
  {Nielsen}, \citenamefont {Severin}, \citenamefont {Backer}, \citenamefont
  {Hyman},\ and\ \citenamefont {Zerial}}]{nielsen1999rab5}%
  \BibitemOpen
  \bibfield  {author} {\bibinfo {author} {\bibfnamefont {E.}~\bibnamefont
  {Nielsen}}, \bibinfo {author} {\bibfnamefont {F.}~\bibnamefont {Severin}},
  \bibinfo {author} {\bibfnamefont {J.~M.}\ \bibnamefont {Backer}}, \bibinfo
  {author} {\bibfnamefont {A.~A.}\ \bibnamefont {Hyman}},\ and\ \bibinfo
  {author} {\bibfnamefont {M.}~\bibnamefont {Zerial}},\ }\bibfield  {title}
  {\bibinfo {title} {Rab5 regulates motility of early endosomes on
  microtubules},\ }\href {https://doi.org/10.1038/14075} {\bibfield  {journal}
  {\bibinfo  {journal} {Nat Cell Biol}\ }\textbf {\bibinfo {volume} {1}},\
  \bibinfo {pages} {376} (\bibinfo {year} {1999})}\BibitemShut {NoStop}%
\bibitem [{\citenamefont {Driskell}\ \emph {et~al.}(2007)\citenamefont
  {Driskell}, \citenamefont {Mironov}, \citenamefont {Allan},\ and\
  \citenamefont {Woodman}}]{driskell2007dynein}%
  \BibitemOpen
  \bibfield  {author} {\bibinfo {author} {\bibfnamefont {O.~J.}\ \bibnamefont
  {Driskell}}, \bibinfo {author} {\bibfnamefont {A.}~\bibnamefont {Mironov}},
  \bibinfo {author} {\bibfnamefont {V.~J.}\ \bibnamefont {Allan}},\ and\
  \bibinfo {author} {\bibfnamefont {P.~G.}\ \bibnamefont {Woodman}},\
  }\bibfield  {title} {\bibinfo {title} {Dynein is required for receptor
  sorting and the morphogenesis of early endosomes},\ }\href
  {https://doi.org/10.1038/ncb1525} {\bibfield  {journal} {\bibinfo  {journal}
  {Nat Cell Biol}\ }\textbf {\bibinfo {volume} {9}},\ \bibinfo {pages} {113}
  (\bibinfo {year} {2007})}\BibitemShut {NoStop}%
\bibitem [{\citenamefont {Flores-Rodriguez}\ \emph {et~al.}(2011)\citenamefont
  {Flores-Rodriguez}, \citenamefont {Rogers}, \citenamefont {Kenwright},
  \citenamefont {Waigh}, \citenamefont {Woodman},\ and\ \citenamefont
  {Allan}}]{flores-rodriguez2011roles}%
  \BibitemOpen
  \bibfield  {author} {\bibinfo {author} {\bibfnamefont {N.}~\bibnamefont
  {Flores-Rodriguez}}, \bibinfo {author} {\bibfnamefont {S.~S.}\ \bibnamefont
  {Rogers}}, \bibinfo {author} {\bibfnamefont {D.~A.}\ \bibnamefont
  {Kenwright}}, \bibinfo {author} {\bibfnamefont {T.~A.}\ \bibnamefont
  {Waigh}}, \bibinfo {author} {\bibfnamefont {P.~G.}\ \bibnamefont {Woodman}},\
  and\ \bibinfo {author} {\bibfnamefont {V.~J.}\ \bibnamefont {Allan}},\
  }\bibfield  {title} {\bibinfo {title} {Roles of {Dynein} and {Dynactin} in
  {Early} {Endosome} {Dynamics} {Revealed} {Using} {Automated} {Tracking} and
  {Global} {Analysis}},\ }\href {https://doi.org/10.1371/journal.pone.0024479}
  {\bibfield  {journal} {\bibinfo  {journal} {PLOS ONE}\ }\textbf {\bibinfo
  {volume} {6}},\ \bibinfo {pages} {e24479} (\bibinfo {year}
  {2011})}\BibitemShut {NoStop}%
\bibitem [{\citenamefont {Misgeld}\ and\ \citenamefont
  {Schwarz}(2017)}]{misgeld2017mitostasis}%
  \BibitemOpen
  \bibfield  {author} {\bibinfo {author} {\bibfnamefont {T.}~\bibnamefont
  {Misgeld}}\ and\ \bibinfo {author} {\bibfnamefont {T.~L.}\ \bibnamefont
  {Schwarz}},\ }\bibfield  {title} {\bibinfo {title} {Mitostasis in neurons:
  maintaining mitochondria in an extended cellular architecture},\ }\href@noop
  {} {\bibfield  {journal} {\bibinfo  {journal} {Neuron}\ }\textbf {\bibinfo
  {volume} {96}},\ \bibinfo {pages} {651} (\bibinfo {year} {2017})}\BibitemShut
  {NoStop}%
\bibitem [{\citenamefont {Presley}\ \emph {et~al.}(1997)\citenamefont
  {Presley}, \citenamefont {Cole}, \citenamefont {Schroer}, \citenamefont
  {Hirschberg}, \citenamefont {Zaal},\ and\ \citenamefont
  {Lippincott-Schwartz}}]{presley1997er}%
  \BibitemOpen
  \bibfield  {author} {\bibinfo {author} {\bibfnamefont {J.~F.}\ \bibnamefont
  {Presley}}, \bibinfo {author} {\bibfnamefont {N.~B.}\ \bibnamefont {Cole}},
  \bibinfo {author} {\bibfnamefont {T.~A.}\ \bibnamefont {Schroer}}, \bibinfo
  {author} {\bibfnamefont {K.}~\bibnamefont {Hirschberg}}, \bibinfo {author}
  {\bibfnamefont {K.~J.}\ \bibnamefont {Zaal}},\ and\ \bibinfo {author}
  {\bibfnamefont {J.}~\bibnamefont {Lippincott-Schwartz}},\ }\bibfield  {title}
  {\bibinfo {title} {Er-to-golgi transport visualized in living cells},\
  }\href@noop {} {\bibfield  {journal} {\bibinfo  {journal} {Nature}\ }\textbf
  {\bibinfo {volume} {389}},\ \bibinfo {pages} {81} (\bibinfo {year}
  {1997})}\BibitemShut {NoStop}%
\bibitem [{\citenamefont {B{\'a}lint}\ \emph {et~al.}(2013)\citenamefont
  {B{\'a}lint}, \citenamefont {Verdeny~Vilanova}, \citenamefont
  {Sandoval~{\'A}lvarez},\ and\ \citenamefont
  {Lakadamyali}}]{balint2013correlative}%
  \BibitemOpen
  \bibfield  {author} {\bibinfo {author} {\bibfnamefont {{\v{S}}.}~\bibnamefont
  {B{\'a}lint}}, \bibinfo {author} {\bibfnamefont {I.}~\bibnamefont
  {Verdeny~Vilanova}}, \bibinfo {author} {\bibfnamefont {{\'A}.}~\bibnamefont
  {Sandoval~{\'A}lvarez}},\ and\ \bibinfo {author} {\bibfnamefont
  {M.}~\bibnamefont {Lakadamyali}},\ }\bibfield  {title} {\bibinfo {title}
  {Correlative live-cell and superresolution microscopy reveals cargo transport
  dynamics at microtubule intersections},\ }\href@noop {} {\bibfield  {journal}
  {\bibinfo  {journal} {P Natl Acad Sci}\ }\textbf {\bibinfo {volume} {110}},\
  \bibinfo {pages} {3375} (\bibinfo {year} {2013})}\BibitemShut {NoStop}%
\bibitem [{\citenamefont {Hancock}(2014)}]{hancock2014bidirectional}%
  \BibitemOpen
  \bibfield  {author} {\bibinfo {author} {\bibfnamefont {W.~O.}\ \bibnamefont
  {Hancock}},\ }\bibfield  {title} {\bibinfo {title} {Bidirectional cargo
  transport: moving beyond tug of war},\ }\href@noop {} {\bibfield  {journal}
  {\bibinfo  {journal} {Nat Rev Mol Cell Bio}\ }\textbf {\bibinfo {volume}
  {15}},\ \bibinfo {pages} {615} (\bibinfo {year} {2014})}\BibitemShut
  {NoStop}%
\bibitem [{\citenamefont {Jongsma}\ \emph {et~al.}(2023)\citenamefont
  {Jongsma}, \citenamefont {Bakker},\ and\ \citenamefont
  {Neefjes}}]{jongsma2023choreographing}%
  \BibitemOpen
  \bibfield  {author} {\bibinfo {author} {\bibfnamefont {M.~L.}\ \bibnamefont
  {Jongsma}}, \bibinfo {author} {\bibfnamefont {N.}~\bibnamefont {Bakker}},\
  and\ \bibinfo {author} {\bibfnamefont {J.}~\bibnamefont {Neefjes}},\
  }\bibfield  {title} {\bibinfo {title} {Choreographing the motor-driven
  endosomal dance},\ }\href@noop {} {\bibfield  {journal} {\bibinfo  {journal}
  {J Cell Sci}\ }\textbf {\bibinfo {volume} {136}},\ \bibinfo {pages}
  {jcs259689} (\bibinfo {year} {2023})}\BibitemShut {NoStop}%
\bibitem [{\citenamefont {Shubeita}\ \emph {et~al.}(2008)\citenamefont
  {Shubeita}, \citenamefont {Tran}, \citenamefont {Xu}, \citenamefont
  {Vershinin}, \citenamefont {Cermelli}, \citenamefont {Cotton}, \citenamefont
  {Welte},\ and\ \citenamefont {Gross}}]{shubeita2008consequences}%
  \BibitemOpen
  \bibfield  {author} {\bibinfo {author} {\bibfnamefont {G.~T.}\ \bibnamefont
  {Shubeita}}, \bibinfo {author} {\bibfnamefont {S.~L.}\ \bibnamefont {Tran}},
  \bibinfo {author} {\bibfnamefont {J.}~\bibnamefont {Xu}}, \bibinfo {author}
  {\bibfnamefont {M.}~\bibnamefont {Vershinin}}, \bibinfo {author}
  {\bibfnamefont {S.}~\bibnamefont {Cermelli}}, \bibinfo {author}
  {\bibfnamefont {S.~L.}\ \bibnamefont {Cotton}}, \bibinfo {author}
  {\bibfnamefont {M.~A.}\ \bibnamefont {Welte}},\ and\ \bibinfo {author}
  {\bibfnamefont {S.~P.}\ \bibnamefont {Gross}},\ }\bibfield  {title} {\bibinfo
  {title} {Consequences of motor copy number on the intracellular transport of
  kinesin-1-driven lipid droplets},\ }\href@noop {} {\bibfield  {journal}
  {\bibinfo  {journal} {Cell}\ }\textbf {\bibinfo {volume} {135}},\ \bibinfo
  {pages} {1098} (\bibinfo {year} {2008})}\BibitemShut {NoStop}%
\bibitem [{\citenamefont {Ahmed}\ \emph {et~al.}(2012)\citenamefont {Ahmed},
  \citenamefont {Li}, \citenamefont {Rubakhin}, \citenamefont {Chiba},
  \citenamefont {Sweedler},\ and\ \citenamefont {Saif}}]{ahmed2012mechanical}%
  \BibitemOpen
  \bibfield  {author} {\bibinfo {author} {\bibfnamefont {W.}~\bibnamefont
  {Ahmed}}, \bibinfo {author} {\bibfnamefont {T.}~\bibnamefont {Li}}, \bibinfo
  {author} {\bibfnamefont {S.}~\bibnamefont {Rubakhin}}, \bibinfo {author}
  {\bibfnamefont {A.}~\bibnamefont {Chiba}}, \bibinfo {author} {\bibfnamefont
  {J.}~\bibnamefont {Sweedler}},\ and\ \bibinfo {author} {\bibfnamefont
  {T.}~\bibnamefont {Saif}},\ }\bibfield  {title} {\bibinfo {title} {Mechanical
  tension modulates local and global vesicle dynamics in neurons},\ }\href@noop
  {} {\bibfield  {journal} {\bibinfo  {journal} {Cell Mol Bioeng}\ }\textbf
  {\bibinfo {volume} {5}},\ \bibinfo {pages} {155} (\bibinfo {year}
  {2012})}\BibitemShut {NoStop}%
\bibitem [{\citenamefont {Higuchi}\ \emph {et~al.}(2014)\citenamefont
  {Higuchi}, \citenamefont {Ashwin}, \citenamefont {Roger},\ and\ \citenamefont
  {Steinberg}}]{higuchi2014early}%
  \BibitemOpen
  \bibfield  {author} {\bibinfo {author} {\bibfnamefont {Y.}~\bibnamefont
  {Higuchi}}, \bibinfo {author} {\bibfnamefont {P.}~\bibnamefont {Ashwin}},
  \bibinfo {author} {\bibfnamefont {Y.}~\bibnamefont {Roger}},\ and\ \bibinfo
  {author} {\bibfnamefont {G.}~\bibnamefont {Steinberg}},\ }\bibfield  {title}
  {\bibinfo {title} {Early endosome motility spatially organizes polysome
  distribution},\ }\href@noop {} {\bibfield  {journal} {\bibinfo  {journal} {J
  Cell Biol}\ }\textbf {\bibinfo {volume} {204}},\ \bibinfo {pages} {343}
  (\bibinfo {year} {2014})}\BibitemShut {NoStop}%
\bibitem [{\citenamefont {Ross}\ \emph {et~al.}(2008)\citenamefont {Ross},
  \citenamefont {Shuman}, \citenamefont {Holzbaur},\ and\ \citenamefont
  {Goldman}}]{ross2008kinesin}%
  \BibitemOpen
  \bibfield  {author} {\bibinfo {author} {\bibfnamefont {J.~L.}\ \bibnamefont
  {Ross}}, \bibinfo {author} {\bibfnamefont {H.}~\bibnamefont {Shuman}},
  \bibinfo {author} {\bibfnamefont {E.~L.~F.}\ \bibnamefont {Holzbaur}},\ and\
  \bibinfo {author} {\bibfnamefont {Y.~E.}\ \bibnamefont {Goldman}},\
  }\bibfield  {title} {\bibinfo {title} {Kinesin and {Dynein}-{Dynactin} at
  {Intersecting} {Microtubules}: {Motor} {Density} {Affects} {Dynein}
  {Function}},\ }\href {https://doi.org/10.1529/biophysj.107.120014} {\bibfield
   {journal} {\bibinfo  {journal} {Biophys J}\ }\textbf {\bibinfo {volume}
  {94}},\ \bibinfo {pages} {3115} (\bibinfo {year} {2008})}\BibitemShut
  {NoStop}%
\bibitem [{\citenamefont {Granger}\ \emph {et~al.}(2014)\citenamefont
  {Granger}, \citenamefont {McNee}, \citenamefont {Allan},\ and\ \citenamefont
  {Woodman}}]{granger2014role}%
  \BibitemOpen
  \bibfield  {author} {\bibinfo {author} {\bibfnamefont {E.}~\bibnamefont
  {Granger}}, \bibinfo {author} {\bibfnamefont {G.}~\bibnamefont {McNee}},
  \bibinfo {author} {\bibfnamefont {V.}~\bibnamefont {Allan}},\ and\ \bibinfo
  {author} {\bibfnamefont {P.}~\bibnamefont {Woodman}},\ }\bibfield  {title}
  {\bibinfo {title} {The role of the cytoskeleton and molecular motors in
  endosomal dynamics},\ }\href {https://doi.org/10.1016/j.semcdb.2014.04.011}
  {\bibfield  {journal} {\bibinfo  {journal} {Semin Cell Dev Biol}\ }\textbf
  {\bibinfo {volume} {31}},\ \bibinfo {pages} {20} (\bibinfo {year}
  {2014})}\BibitemShut {NoStop}%
\bibitem [{\citenamefont {Mogre}\ and\ \citenamefont
  {Koslover}(2018)}]{mogre2018multimodal}%
  \BibitemOpen
  \bibfield  {author} {\bibinfo {author} {\bibfnamefont {S.~S.}\ \bibnamefont
  {Mogre}}\ and\ \bibinfo {author} {\bibfnamefont {E.~F.}\ \bibnamefont
  {Koslover}},\ }\bibfield  {title} {\bibinfo {title} {Multimodal transport and
  dispersion of organelles in narrow tubular cells},\ }\href@noop {} {\bibfield
   {journal} {\bibinfo  {journal} {Phys Rev E}\ }\textbf {\bibinfo {volume}
  {97}},\ \bibinfo {pages} {042402} (\bibinfo {year} {2018})}\BibitemShut
  {NoStop}%
\bibitem [{\citenamefont {Hafner}\ and\ \citenamefont
  {Rieger}(2018)}]{hafner2018spatial}%
  \BibitemOpen
  \bibfield  {author} {\bibinfo {author} {\bibfnamefont {A.~E.}\ \bibnamefont
  {Hafner}}\ and\ \bibinfo {author} {\bibfnamefont {H.}~\bibnamefont
  {Rieger}},\ }\bibfield  {title} {\bibinfo {title} {Spatial cytoskeleton
  organization supports targeted intracellular transport},\ }\href@noop {}
  {\bibfield  {journal} {\bibinfo  {journal} {Biophys J}\ }\textbf {\bibinfo
  {volume} {114}},\ \bibinfo {pages} {1420} (\bibinfo {year}
  {2018})}\BibitemShut {NoStop}%
\bibitem [{\citenamefont {Hafner}\ and\ \citenamefont
  {Rieger}(2016)}]{hafner2016spatial}%
  \BibitemOpen
  \bibfield  {author} {\bibinfo {author} {\bibfnamefont {A.~E.}\ \bibnamefont
  {Hafner}}\ and\ \bibinfo {author} {\bibfnamefont {H.}~\bibnamefont
  {Rieger}},\ }\bibfield  {title} {\bibinfo {title} {Spatial organization of
  the cytoskeleton enhances cargo delivery to specific target areas on the
  plasma membrane of spherical cells},\ }\bibfield  {journal} {\bibinfo
  {journal} {Phys Biol}\ }\textbf {\bibinfo {volume} {13}},\ \href
  {https://doi.org/10.1088/1478-3975/13/6/066003}
  {10.1088/1478-3975/13/6/066003} (\bibinfo {year} {2016})\BibitemShut
  {NoStop}%
\bibitem [{\citenamefont {Ando}\ \emph {et~al.}(2015)\citenamefont {Ando},
  \citenamefont {Korabel}, \citenamefont {Huang},\ and\ \citenamefont
  {Gopinathan}}]{ando2015cytoskeletal}%
  \BibitemOpen
  \bibfield  {author} {\bibinfo {author} {\bibfnamefont {D.}~\bibnamefont
  {Ando}}, \bibinfo {author} {\bibfnamefont {N.}~\bibnamefont {Korabel}},
  \bibinfo {author} {\bibfnamefont {K.~C.}\ \bibnamefont {Huang}},\ and\
  \bibinfo {author} {\bibfnamefont {A.}~\bibnamefont {Gopinathan}},\ }\bibfield
   {title} {\bibinfo {title} {Cytoskeletal network morphology regulates
  intracellular transport dynamics},\ }\href@noop {} {\bibfield  {journal}
  {\bibinfo  {journal} {Biophys J}\ }\textbf {\bibinfo {volume} {109}},\
  \bibinfo {pages} {1574} (\bibinfo {year} {2015})}\BibitemShut {NoStop}%
\bibitem [{\citenamefont {Campos}\ \emph {et~al.}(2015)\citenamefont {Campos},
  \citenamefont {Abad}, \citenamefont {M{\'e}ndez}, \citenamefont {Yuste},\
  and\ \citenamefont {Lindenberg}}]{campos2015optimal}%
  \BibitemOpen
  \bibfield  {author} {\bibinfo {author} {\bibfnamefont {D.}~\bibnamefont
  {Campos}}, \bibinfo {author} {\bibfnamefont {E.}~\bibnamefont {Abad}},
  \bibinfo {author} {\bibfnamefont {V.}~\bibnamefont {M{\'e}ndez}}, \bibinfo
  {author} {\bibfnamefont {S.}~\bibnamefont {Yuste}},\ and\ \bibinfo {author}
  {\bibfnamefont {K.}~\bibnamefont {Lindenberg}},\ }\bibfield  {title}
  {\bibinfo {title} {Optimal search strategies of space-time coupled random
  walkers with finite lifetimes},\ }\href@noop {} {\bibfield  {journal}
  {\bibinfo  {journal} {Phys Rev E}\ }\textbf {\bibinfo {volume} {91}},\
  \bibinfo {pages} {052115} (\bibinfo {year} {2015})}\BibitemShut {NoStop}%
\bibitem [{\citenamefont {Hafner}\ \emph {et~al.}(2016)\citenamefont {Hafner},
  \citenamefont {Santen}, \citenamefont {Rieger},\ and\ \citenamefont
  {Shaebani}}]{hafner2016run}%
  \BibitemOpen
  \bibfield  {author} {\bibinfo {author} {\bibfnamefont {A.~E.}\ \bibnamefont
  {Hafner}}, \bibinfo {author} {\bibfnamefont {L.}~\bibnamefont {Santen}},
  \bibinfo {author} {\bibfnamefont {H.}~\bibnamefont {Rieger}},\ and\ \bibinfo
  {author} {\bibfnamefont {M.~R.}\ \bibnamefont {Shaebani}},\ }\bibfield
  {title} {\bibinfo {title} {Run-and-pause dynamics of cytoskeletal motor
  proteins},\ }\href@noop {} {\bibfield  {journal} {\bibinfo  {journal} {Sci
  Rep}\ }\textbf {\bibinfo {volume} {6}},\ \bibinfo {pages} {37162} (\bibinfo
  {year} {2016})}\BibitemShut {NoStop}%
\bibitem [{\citenamefont {Bénichou}\ \emph {et~al.}(2011)\citenamefont
  {Bénichou}, \citenamefont {Loverdo}, \citenamefont {Moreau},\ and\
  \citenamefont {Voituriez}}]{benichou2011intermittent}%
  \BibitemOpen
  \bibfield  {author} {\bibinfo {author} {\bibfnamefont {O.}~\bibnamefont
  {Bénichou}}, \bibinfo {author} {\bibfnamefont {C.}~\bibnamefont {Loverdo}},
  \bibinfo {author} {\bibfnamefont {M.}~\bibnamefont {Moreau}},\ and\ \bibinfo
  {author} {\bibfnamefont {R.}~\bibnamefont {Voituriez}},\ }\bibfield  {title}
  {\bibinfo {title} {Intermittent search strategies},\ }\href
  {https://doi.org/10.1103/RevModPhys.83.81} {\bibfield  {journal} {\bibinfo
  {journal} {Rev Mod Phys}\ }\textbf {\bibinfo {volume} {83}},\ \bibinfo
  {pages} {81} (\bibinfo {year} {2011})}\BibitemShut {NoStop}%
\bibitem [{\citenamefont {Le~Vot}\ \emph {et~al.}(2022)\citenamefont {Le~Vot},
  \citenamefont {Yuste}, \citenamefont {Abad},\ and\ \citenamefont
  {Grebenkov}}]{le_vot2022first}%
  \BibitemOpen
  \bibfield  {author} {\bibinfo {author} {\bibfnamefont {F.}~\bibnamefont
  {Le~Vot}}, \bibinfo {author} {\bibfnamefont {S.~B.}\ \bibnamefont {Yuste}},
  \bibinfo {author} {\bibfnamefont {E.}~\bibnamefont {Abad}},\ and\ \bibinfo
  {author} {\bibfnamefont {D.~S.}\ \bibnamefont {Grebenkov}},\ }\bibfield
  {title} {\bibinfo {title} {First-encounter time of two diffusing particles in
  two- and three-dimensional confinement},\ }\href
  {https://doi.org/10.1103/PhysRevE.105.044119} {\bibfield  {journal} {\bibinfo
   {journal} {Phys Rev E}\ }\textbf {\bibinfo {volume} {105}},\ \bibinfo
  {pages} {044119} (\bibinfo {year} {2022})}\BibitemShut {NoStop}%
\bibitem [{\citenamefont {Rupprecht}\ \emph {et~al.}(2016)\citenamefont
  {Rupprecht}, \citenamefont {Bénichou},\ and\ \citenamefont
  {Voituriez}}]{rupprecht2016optimal}%
  \BibitemOpen
  \bibfield  {author} {\bibinfo {author} {\bibfnamefont {J.-F.}\ \bibnamefont
  {Rupprecht}}, \bibinfo {author} {\bibfnamefont {O.}~\bibnamefont
  {Bénichou}},\ and\ \bibinfo {author} {\bibfnamefont {R.}~\bibnamefont
  {Voituriez}},\ }\bibfield  {title} {\bibinfo {title} {Optimal search
  strategies of run-and-tumble walks},\ }\href
  {https://doi.org/10.1103/PhysRevE.94.012117} {\bibfield  {journal} {\bibinfo
  {journal} {Phys Rev E}\ }\textbf {\bibinfo {volume} {94}},\ \bibinfo {pages}
  {012117} (\bibinfo {year} {2016})}\BibitemShut {NoStop}%
\bibitem [{\citenamefont {Maelfeyt}\ \emph {et~al.}(2019)\citenamefont
  {Maelfeyt}, \citenamefont {Tabei},\ and\ \citenamefont
  {Gopinathan}}]{maelfeyt2019anomalous}%
  \BibitemOpen
  \bibfield  {author} {\bibinfo {author} {\bibfnamefont {B.}~\bibnamefont
  {Maelfeyt}}, \bibinfo {author} {\bibfnamefont {S.~M.~A.}\ \bibnamefont
  {Tabei}},\ and\ \bibinfo {author} {\bibfnamefont {A.}~\bibnamefont
  {Gopinathan}},\ }\bibfield  {title} {\bibinfo {title} {Anomalous
  intracellular transport phases depend on cytoskeletal network features},\
  }\bibfield  {journal} {\bibinfo  {journal} {Phys Rev E}\ }\textbf {\bibinfo
  {volume} {99}},\ \href {https://doi.org/10.1103/PhysRevE.99.062404}
  {10.1103/PhysRevE.99.062404} (\bibinfo {year} {2019})\BibitemShut {NoStop}%
\bibitem [{\citenamefont {Mlynarczyk}\ and\ \citenamefont
  {Abel}(2019)}]{mlynarczyk2019first}%
  \BibitemOpen
  \bibfield  {author} {\bibinfo {author} {\bibfnamefont {P.~J.}\ \bibnamefont
  {Mlynarczyk}}\ and\ \bibinfo {author} {\bibfnamefont {S.~M.}\ \bibnamefont
  {Abel}},\ }\bibfield  {title} {\bibinfo {title} {First passage of molecular
  motors on networks of cytoskeletal filaments},\ }\bibfield  {journal}
  {\bibinfo  {journal} {Phys Rev E}\ }\textbf {\bibinfo {volume} {99}},\ \href
  {https://doi.org/10.1103/PhysRevE.99.022406} {10.1103/PhysRevE.99.022406}
  (\bibinfo {year} {2019})\BibitemShut {NoStop}%
\bibitem [{\citenamefont {Ciocanel}\ \emph {et~al.}(2018)\citenamefont
  {Ciocanel}, \citenamefont {Sandstede}, \citenamefont {Jeschonek},\ and\
  \citenamefont {Mowry}}]{ciocanel2018modeling}%
  \BibitemOpen
  \bibfield  {author} {\bibinfo {author} {\bibfnamefont {M.-V.}\ \bibnamefont
  {Ciocanel}}, \bibinfo {author} {\bibfnamefont {B.}~\bibnamefont {Sandstede}},
  \bibinfo {author} {\bibfnamefont {S.~P.}\ \bibnamefont {Jeschonek}},\ and\
  \bibinfo {author} {\bibfnamefont {K.~L.}\ \bibnamefont {Mowry}},\ }\bibfield
  {title} {\bibinfo {title} {Modeling microtubule-based transport and anchoring
  of mrna},\ }\href@noop {} {\bibfield  {journal} {\bibinfo  {journal} {Siam J
  Appl Dyn Syst}\ }\textbf {\bibinfo {volume} {17}},\ \bibinfo {pages} {2855}
  (\bibinfo {year} {2018})}\BibitemShut {NoStop}%
\bibitem [{\citenamefont {{Tejedor}}\ \emph {et~al.}(2011)\citenamefont
  {{Tejedor}}, \citenamefont {{Schad}}, \citenamefont {{Metzler}},
  \citenamefont {{Benichou}},\ and\ \citenamefont
  {{Voituriez}}}]{tejedor2011encounter}%
  \BibitemOpen
  \bibfield  {author} {\bibinfo {author} {\bibfnamefont {V.}~\bibnamefont
  {{Tejedor}}}, \bibinfo {author} {\bibfnamefont {M.}~\bibnamefont {{Schad}}},
  \bibinfo {author} {\bibfnamefont {R.}~\bibnamefont {{Metzler}}}, \bibinfo
  {author} {\bibfnamefont {O.}~\bibnamefont {{Benichou}}},\ and\ \bibinfo
  {author} {\bibfnamefont {R.}~\bibnamefont {{Voituriez}}},\ }\bibfield
  {title} {\bibinfo {title} {Encounter distribution of two random walkers on a
  finite one-dimensional interval},\ }\href@noop {} {\bibfield  {journal}
  {\bibinfo  {journal} {J Phys A: Math Theor}\ }\textbf {\bibinfo {volume}
  {44}} (\bibinfo {year} {2011})}\BibitemShut {NoStop}%
\bibitem [{\citenamefont {Scholz}\ \emph {et~al.}(2016)\citenamefont {Scholz},
  \citenamefont {Burov}, \citenamefont {Weirich}, \citenamefont {Scholz},
  \citenamefont {Tabei}, \citenamefont {Gardel},\ and\ \citenamefont
  {Dinner}}]{scholz2016cycling}%
  \BibitemOpen
  \bibfield  {author} {\bibinfo {author} {\bibfnamefont {M.}~\bibnamefont
  {Scholz}}, \bibinfo {author} {\bibfnamefont {S.}~\bibnamefont {Burov}},
  \bibinfo {author} {\bibfnamefont {K.~L.}\ \bibnamefont {Weirich}}, \bibinfo
  {author} {\bibfnamefont {B.~J.}\ \bibnamefont {Scholz}}, \bibinfo {author}
  {\bibfnamefont {S.~A.}\ \bibnamefont {Tabei}}, \bibinfo {author}
  {\bibfnamefont {M.~L.}\ \bibnamefont {Gardel}},\ and\ \bibinfo {author}
  {\bibfnamefont {A.~R.}\ \bibnamefont {Dinner}},\ }\bibfield  {title}
  {\bibinfo {title} {Cycling state that can lead to glassy dynamics in
  intracellular transport},\ }\href@noop {} {\bibfield  {journal} {\bibinfo
  {journal} {Phys Rev X}\ }\textbf {\bibinfo {volume} {6}},\ \bibinfo {pages}
  {011037} (\bibinfo {year} {2016})}\BibitemShut {NoStop}%
\bibitem [{\citenamefont {Schwarz}\ \emph {et~al.}(2016)\citenamefont
  {Schwarz}, \citenamefont {Schr{\"o}der}, \citenamefont {Qu}, \citenamefont
  {Hoth},\ and\ \citenamefont {Rieger}}]{schwarz2016optimality}%
  \BibitemOpen
  \bibfield  {author} {\bibinfo {author} {\bibfnamefont {K.}~\bibnamefont
  {Schwarz}}, \bibinfo {author} {\bibfnamefont {Y.}~\bibnamefont
  {Schr{\"o}der}}, \bibinfo {author} {\bibfnamefont {B.}~\bibnamefont {Qu}},
  \bibinfo {author} {\bibfnamefont {M.}~\bibnamefont {Hoth}},\ and\ \bibinfo
  {author} {\bibfnamefont {H.}~\bibnamefont {Rieger}},\ }\bibfield  {title}
  {\bibinfo {title} {Optimality of spatially inhomogeneous search strategies},\
  }\href@noop {} {\bibfield  {journal} {\bibinfo  {journal} {Phys Rev Lett}\
  }\textbf {\bibinfo {volume} {117}},\ \bibinfo {pages} {068101} (\bibinfo
  {year} {2016})}\BibitemShut {NoStop}%
\bibitem [{\citenamefont {Hafner}\ and\ \citenamefont
  {Rieger}(2020)}]{hafner2020spatially}%
  \BibitemOpen
  \bibfield  {author} {\bibinfo {author} {\bibfnamefont {A.}~\bibnamefont
  {Hafner}}\ and\ \bibinfo {author} {\bibfnamefont {H.}~\bibnamefont
  {Rieger}},\ }\bibfield  {title} {\bibinfo {title} {Spatially inhomogeneous
  search strategies},\ }in\ \href@noop {} {\emph {\bibinfo {booktitle}
  {Chemical Kinetics: Beyond the Textbook}}}\ (\bibinfo  {publisher} {World
  Scientific},\ \bibinfo {year} {2020})\ pp.\ \bibinfo {pages}
  {285--302}\BibitemShut {NoStop}%
\bibitem [{\citenamefont {Scott}\ \emph {et~al.}(2021)\citenamefont {Scott},
  \citenamefont {Brown}, \citenamefont {Mogre}, \citenamefont {Westrate},\ and\
  \citenamefont {Koslover}}]{scott2021diffusive}%
  \BibitemOpen
  \bibfield  {author} {\bibinfo {author} {\bibfnamefont {Z.~C.}\ \bibnamefont
  {Scott}}, \bibinfo {author} {\bibfnamefont {A.~I.}\ \bibnamefont {Brown}},
  \bibinfo {author} {\bibfnamefont {S.~S.}\ \bibnamefont {Mogre}}, \bibinfo
  {author} {\bibfnamefont {L.~M.}\ \bibnamefont {Westrate}},\ and\ \bibinfo
  {author} {\bibfnamefont {E.~F.}\ \bibnamefont {Koslover}},\ }\bibfield
  {title} {\bibinfo {title} {Diffusive search and trajectories on tubular
  networks: a propagator approach},\ }\href@noop {} {\bibfield  {journal}
  {\bibinfo  {journal} {Eur Phys J E}\ }\textbf {\bibinfo {volume} {44}},\
  \bibinfo {pages} {80} (\bibinfo {year} {2021})}\BibitemShut {NoStop}%
\bibitem [{\citenamefont {Scott}\ \emph {et~al.}(2023)\citenamefont {Scott},
  \citenamefont {Koning}, \citenamefont {Vanderwerp}, \citenamefont {Cohen},
  \citenamefont {Westrate},\ and\ \citenamefont
  {Koslover}}]{scott2023endoplasmic}%
  \BibitemOpen
  \bibfield  {author} {\bibinfo {author} {\bibfnamefont {Z.~C.}\ \bibnamefont
  {Scott}}, \bibinfo {author} {\bibfnamefont {K.}~\bibnamefont {Koning}},
  \bibinfo {author} {\bibfnamefont {M.}~\bibnamefont {Vanderwerp}}, \bibinfo
  {author} {\bibfnamefont {L.}~\bibnamefont {Cohen}}, \bibinfo {author}
  {\bibfnamefont {L.~M.}\ \bibnamefont {Westrate}},\ and\ \bibinfo {author}
  {\bibfnamefont {E.~F.}\ \bibnamefont {Koslover}},\ }\bibfield  {title}
  {\bibinfo {title} {Endoplasmic reticulum network heterogeneity guides
  diffusive transport and kinetics},\ }\href@noop {} {\bibfield  {journal}
  {\bibinfo  {journal} {Biophys J}\ }\textbf {\bibinfo {volume} {122}}
  (\bibinfo {year} {2023})}\BibitemShut {NoStop}%
\bibitem [{\citenamefont {Mori}\ \emph {et~al.}(2020)\citenamefont {Mori},
  \citenamefont {Le~Doussal}, \citenamefont {Majumdar},\ and\ \citenamefont
  {Schehr}}]{mori2020universal}%
  \BibitemOpen
  \bibfield  {author} {\bibinfo {author} {\bibfnamefont {F.}~\bibnamefont
  {Mori}}, \bibinfo {author} {\bibfnamefont {P.}~\bibnamefont {Le~Doussal}},
  \bibinfo {author} {\bibfnamefont {S.~N.}\ \bibnamefont {Majumdar}},\ and\
  \bibinfo {author} {\bibfnamefont {G.}~\bibnamefont {Schehr}},\ }\bibfield
  {title} {\bibinfo {title} {Universal survival probability for a
  $d$-dimensional run-and-tumble particle},\ }\bibfield  {journal} {\bibinfo
  {journal} {Phys Rev Lett}\ }\textbf {\bibinfo {volume} {124}},\ \href
  {https://doi.org/10.1103/physrevlett.124.090603}
  {10.1103/physrevlett.124.090603} (\bibinfo {year} {2020})\BibitemShut
  {NoStop}%
\bibitem [{\citenamefont {Bartumeus}\ \emph {et~al.}(2002)\citenamefont
  {Bartumeus}, \citenamefont {Catalan}, \citenamefont {Fulco}, \citenamefont
  {Lyra},\ and\ \citenamefont {Viswanathan}}]{bartumeus2002optimizing}%
  \BibitemOpen
  \bibfield  {author} {\bibinfo {author} {\bibfnamefont {F.}~\bibnamefont
  {Bartumeus}}, \bibinfo {author} {\bibfnamefont {J.}~\bibnamefont {Catalan}},
  \bibinfo {author} {\bibfnamefont {U.~L.}\ \bibnamefont {Fulco}}, \bibinfo
  {author} {\bibfnamefont {M.~L.}\ \bibnamefont {Lyra}},\ and\ \bibinfo
  {author} {\bibfnamefont {G.~M.}\ \bibnamefont {Viswanathan}},\ }\bibfield
  {title} {\bibinfo {title} {Optimizing the encounter rate in biological
  interactions: L\'evy versus brownian strategies},\ }\href
  {https://doi.org/10.1103/PhysRevLett.88.097901} {\bibfield  {journal}
  {\bibinfo  {journal} {Phys Rev Lett}\ }\textbf {\bibinfo {volume} {88}},\
  \bibinfo {pages} {097901} (\bibinfo {year} {2002})}\BibitemShut {NoStop}%
\bibitem [{\citenamefont {Loverdo}\ \emph {et~al.}(2008)\citenamefont
  {Loverdo}, \citenamefont {Bénichou}, \citenamefont {Moreau},\ and\
  \citenamefont {Voituriez}}]{loverdo2008enhanced}%
  \BibitemOpen
  \bibfield  {author} {\bibinfo {author} {\bibfnamefont {C.}~\bibnamefont
  {Loverdo}}, \bibinfo {author} {\bibfnamefont {O.}~\bibnamefont {Bénichou}},
  \bibinfo {author} {\bibfnamefont {M.}~\bibnamefont {Moreau}},\ and\ \bibinfo
  {author} {\bibfnamefont {R.}~\bibnamefont {Voituriez}},\ }\bibfield  {title}
  {\bibinfo {title} {Enhanced reaction kinetics in biological cells},\ }\href
  {https://doi.org/10.1038/nphys830} {\bibfield  {journal} {\bibinfo  {journal}
  {Nat Phys}\ }\textbf {\bibinfo {volume} {4}},\ \bibinfo {pages} {134–137}
  (\bibinfo {year} {2008})}\BibitemShut {NoStop}%
\bibitem [{\citenamefont {Pangarkar}\ \emph {et~al.}(2005)\citenamefont
  {Pangarkar}, \citenamefont {Dinh},\ and\ \citenamefont
  {Mitragotri}}]{pangarkar2005dynamics}%
  \BibitemOpen
  \bibfield  {author} {\bibinfo {author} {\bibfnamefont {C.}~\bibnamefont
  {Pangarkar}}, \bibinfo {author} {\bibfnamefont {A.~T.}\ \bibnamefont
  {Dinh}},\ and\ \bibinfo {author} {\bibfnamefont {S.}~\bibnamefont
  {Mitragotri}},\ }\bibfield  {title} {\bibinfo {title} {Dynamics and {Spatial}
  {Organization} of {Endosomes} in {Mammalian} {Cells}},\ }\href
  {https://doi.org/10.1103/PhysRevLett.95.158101} {\bibfield  {journal}
  {\bibinfo  {journal} {Phys Rev Lett}\ }\textbf {\bibinfo {volume} {95}},\
  \bibinfo {pages} {158101} (\bibinfo {year} {2005})},\ \bibinfo {note}
  {publisher: American Physical Society}\BibitemShut {NoStop}%
\bibitem [{\citenamefont {Berg}(1983)}]{berg1983random}%
  \BibitemOpen
  \bibfield  {author} {\bibinfo {author} {\bibfnamefont {H.~C.}\ \bibnamefont
  {Berg}},\ }\href@noop {} {\emph {\bibinfo {title} {Random Walks in
  Biology}}}\ (\bibinfo  {publisher} {Princeton University Press},\ \bibinfo
  {address} {Princeton, NJ},\ \bibinfo {year} {1983})\BibitemShut {NoStop}%
\bibitem [{\citenamefont {Kahana}\ \emph {et~al.}(2008)\citenamefont {Kahana},
  \citenamefont {Kenan}, \citenamefont {Feingold}, \citenamefont {Elbaum},\
  and\ \citenamefont {Granek}}]{kahana2008active}%
  \BibitemOpen
  \bibfield  {author} {\bibinfo {author} {\bibfnamefont {A.}~\bibnamefont
  {Kahana}}, \bibinfo {author} {\bibfnamefont {G.}~\bibnamefont {Kenan}},
  \bibinfo {author} {\bibfnamefont {M.}~\bibnamefont {Feingold}}, \bibinfo
  {author} {\bibfnamefont {M.}~\bibnamefont {Elbaum}},\ and\ \bibinfo {author}
  {\bibfnamefont {R.}~\bibnamefont {Granek}},\ }\bibfield  {title} {\bibinfo
  {title} {Active transport on disordered microtubule networks: The generalized
  random velocity model},\ }\href@noop {} {\bibfield  {journal} {\bibinfo
  {journal} {Phys Rev E}\ }\textbf {\bibinfo {volume} {78}},\ \bibinfo {pages}
  {051912} (\bibinfo {year} {2008})}\BibitemShut {NoStop}%
\bibitem [{\citenamefont {Verkman}(2002)}]{verkman2002solute}%
  \BibitemOpen
  \bibfield  {author} {\bibinfo {author} {\bibfnamefont {A.~S.}\ \bibnamefont
  {Verkman}},\ }\bibfield  {title} {\bibinfo {title} {Solute and macromolecule
  diffusion in cellular aqueous compartments},\ }\href@noop {} {\bibfield
  {journal} {\bibinfo  {journal} {Trends Biochem Sci}\ }\textbf {\bibinfo
  {volume} {27}},\ \bibinfo {pages} {27} (\bibinfo {year} {2002})}\BibitemShut
  {NoStop}%
\bibitem [{\citenamefont {Koslover}\ \emph {et~al.}(2025)\citenamefont
  {Koslover}, \citenamefont {Lin},\ and\ \citenamefont
  {Phillips}}]{koslover2025many}%
  \BibitemOpen
  \bibfield  {author} {\bibinfo {author} {\bibfnamefont {E.~F.}\ \bibnamefont
  {Koslover}}, \bibinfo {author} {\bibfnamefont {M.~M.}\ \bibnamefont {Lin}},\
  and\ \bibinfo {author} {\bibfnamefont {R.}~\bibnamefont {Phillips}},\
  }\bibfield  {title} {\bibinfo {title} {Many will enter, few will win: Cost
  and sensitivity of exploratory dynamics},\ }\href@noop {} {\bibfield
  {journal} {\bibinfo  {journal} {arXiv preprint arXiv:2506.00775}\ } (\bibinfo
  {year} {2025})}\BibitemShut {NoStop}%
\bibitem [{\citenamefont {Guimaraes}\ \emph {et~al.}(2015)\citenamefont
  {Guimaraes}, \citenamefont {Schuster}, \citenamefont {Bielska}, \citenamefont
  {Dagdas}, \citenamefont {Kilaru}, \citenamefont {Meadows}, \citenamefont
  {Schrader},\ and\ \citenamefont {Steinberg}}]{guimaraes2015peroxisomes}%
  \BibitemOpen
  \bibfield  {author} {\bibinfo {author} {\bibfnamefont {S.~C.}\ \bibnamefont
  {Guimaraes}}, \bibinfo {author} {\bibfnamefont {M.}~\bibnamefont {Schuster}},
  \bibinfo {author} {\bibfnamefont {E.}~\bibnamefont {Bielska}}, \bibinfo
  {author} {\bibfnamefont {G.}~\bibnamefont {Dagdas}}, \bibinfo {author}
  {\bibfnamefont {S.}~\bibnamefont {Kilaru}}, \bibinfo {author} {\bibfnamefont
  {B.~R.}\ \bibnamefont {Meadows}}, \bibinfo {author} {\bibfnamefont
  {M.}~\bibnamefont {Schrader}},\ and\ \bibinfo {author} {\bibfnamefont
  {G.}~\bibnamefont {Steinberg}},\ }\bibfield  {title} {\bibinfo {title}
  {Peroxisomes, lipid droplets, and endoplasmic reticulum “hitchhike” on
  motile early endosomes},\ }\href@noop {} {\bibfield  {journal} {\bibinfo
  {journal} {J Cell Biol}\ }\textbf {\bibinfo {volume} {211}},\ \bibinfo
  {pages} {945} (\bibinfo {year} {2015})}\BibitemShut {NoStop}%
\bibitem [{\citenamefont {{Weng}}\ \emph {et~al.}(2017)\citenamefont {{Weng}},
  \citenamefont {{Zhang}}, \citenamefont {{Small}},\ and\ \citenamefont
  {{Hui}}}]{weng2017hunting}%
  \BibitemOpen
  \bibfield  {author} {\bibinfo {author} {\bibfnamefont {T.}~\bibnamefont
  {{Weng}}}, \bibinfo {author} {\bibfnamefont {J.}~\bibnamefont {{Zhang}}},
  \bibinfo {author} {\bibfnamefont {M.}~\bibnamefont {{Small}}},\ and\ \bibinfo
  {author} {\bibfnamefont {P.}~\bibnamefont {{Hui}}},\ }\bibfield  {title}
  {\bibinfo {title} {Hunting for a moving target on a complex network},\ }\href
  {https://doi.org/10.1209/0295-5075/119/48006} {\bibfield  {journal} {\bibinfo
   {journal} {Europhys Lett}\ }\textbf {\bibinfo {volume} {119}},\ \bibinfo
  {pages} {48006} (\bibinfo {year} {2017})}\BibitemShut {NoStop}%
\bibitem [{\citenamefont {Burute}\ and\ \citenamefont
  {Kapitein}(2019)}]{burute2019cellular}%
  \BibitemOpen
  \bibfield  {author} {\bibinfo {author} {\bibfnamefont {M.}~\bibnamefont
  {Burute}}\ and\ \bibinfo {author} {\bibfnamefont {L.~C.}\ \bibnamefont
  {Kapitein}},\ }\bibfield  {title} {\bibinfo {title} {Cellular logistics:
  unraveling the interplay between microtubule organization and intracellular
  transport},\ }\href@noop {} {\bibfield  {journal} {\bibinfo  {journal} {Annu
  Rev Cell Dev Bi}\ }\textbf {\bibinfo {volume} {35}},\ \bibinfo {pages} {29}
  (\bibinfo {year} {2019})}\BibitemShut {NoStop}%
\bibitem [{\citenamefont {Sallee}\ and\ \citenamefont
  {Feldman}(2021)}]{sallee2021microtubule}%
  \BibitemOpen
  \bibfield  {author} {\bibinfo {author} {\bibfnamefont {M.~D.}\ \bibnamefont
  {Sallee}}\ and\ \bibinfo {author} {\bibfnamefont {J.~L.}\ \bibnamefont
  {Feldman}},\ }\bibfield  {title} {\bibinfo {title} {Microtubule organization
  across cell types and states},\ }\href@noop {} {\bibfield  {journal}
  {\bibinfo  {journal} {Curr Biol}\ }\textbf {\bibinfo {volume} {31}},\
  \bibinfo {pages} {R506} (\bibinfo {year} {2021})}\BibitemShut {NoStop}%
\bibitem [{\citenamefont {Skau}\ and\ \citenamefont
  {Waterman}(2015)}]{skau2015specification}%
  \BibitemOpen
  \bibfield  {author} {\bibinfo {author} {\bibfnamefont {C.~T.}\ \bibnamefont
  {Skau}}\ and\ \bibinfo {author} {\bibfnamefont {C.~M.}\ \bibnamefont
  {Waterman}},\ }\bibfield  {title} {\bibinfo {title} {Specification of
  architecture and function of actin structures by actin nucleation factors},\
  }\href@noop {} {\bibfield  {journal} {\bibinfo  {journal} {Ann Rev Biophys}\
  }\textbf {\bibinfo {volume} {44}},\ \bibinfo {pages} {285} (\bibinfo {year}
  {2015})}\BibitemShut {NoStop}%
\bibitem [{\citenamefont {Oberhofer}\ \emph {et~al.}(2020)\citenamefont
  {Oberhofer}, \citenamefont {Reithmann}, \citenamefont {Spieler},
  \citenamefont {Stepp}, \citenamefont {Zimmermann}, \citenamefont {Schmid},
  \citenamefont {Frey},\ and\ \citenamefont
  {{\"O}kten}}]{oberhofer2020molecular}%
  \BibitemOpen
  \bibfield  {author} {\bibinfo {author} {\bibfnamefont {A.}~\bibnamefont
  {Oberhofer}}, \bibinfo {author} {\bibfnamefont {E.}~\bibnamefont
  {Reithmann}}, \bibinfo {author} {\bibfnamefont {P.}~\bibnamefont {Spieler}},
  \bibinfo {author} {\bibfnamefont {W.~L.}\ \bibnamefont {Stepp}}, \bibinfo
  {author} {\bibfnamefont {D.}~\bibnamefont {Zimmermann}}, \bibinfo {author}
  {\bibfnamefont {B.}~\bibnamefont {Schmid}}, \bibinfo {author} {\bibfnamefont
  {E.}~\bibnamefont {Frey}},\ and\ \bibinfo {author} {\bibfnamefont
  {Z.}~\bibnamefont {{\"O}kten}},\ }\bibfield  {title} {\bibinfo {title}
  {Molecular underpinnings of cytoskeletal cross-talk},\ }\href@noop {}
  {\bibfield  {journal} {\bibinfo  {journal} {P Natl Acad Sci}\ }\textbf
  {\bibinfo {volume} {117}},\ \bibinfo {pages} {3944} (\bibinfo {year}
  {2020})}\BibitemShut {NoStop}%
\bibitem [{\citenamefont {Broedersz}\ and\ \citenamefont
  {MacKintosh}(2014)}]{broedersz2014modeling}%
  \BibitemOpen
  \bibfield  {author} {\bibinfo {author} {\bibfnamefont {C.~P.}\ \bibnamefont
  {Broedersz}}\ and\ \bibinfo {author} {\bibfnamefont {F.~C.}\ \bibnamefont
  {MacKintosh}},\ }\bibfield  {title} {\bibinfo {title} {Modeling semiflexible
  polymer networks},\ }\href@noop {} {\bibfield  {journal} {\bibinfo  {journal}
  {Rev Mod Phys}\ }\textbf {\bibinfo {volume} {86}},\ \bibinfo {pages} {995}
  (\bibinfo {year} {2014})}\BibitemShut {NoStop}%
\bibitem [{\citenamefont {Burakov}\ \emph {et~al.}(2021)\citenamefont
  {Burakov}, \citenamefont {Vorobjev}, \citenamefont {Semenova}, \citenamefont
  {Cowan}, \citenamefont {Carson}, \citenamefont {Wu},\ and\ \citenamefont
  {Rodionov}}]{burakov2021persistent}%
  \BibitemOpen
  \bibfield  {author} {\bibinfo {author} {\bibfnamefont {A.}~\bibnamefont
  {Burakov}}, \bibinfo {author} {\bibfnamefont {I.}~\bibnamefont {Vorobjev}},
  \bibinfo {author} {\bibfnamefont {I.}~\bibnamefont {Semenova}}, \bibinfo
  {author} {\bibfnamefont {A.}~\bibnamefont {Cowan}}, \bibinfo {author}
  {\bibfnamefont {J.}~\bibnamefont {Carson}}, \bibinfo {author} {\bibfnamefont
  {Y.}~\bibnamefont {Wu}},\ and\ \bibinfo {author} {\bibfnamefont
  {V.}~\bibnamefont {Rodionov}},\ }\bibfield  {title} {\bibinfo {title}
  {Persistent growth of microtubules at low density},\ }\href
  {https://doi.org/10.1091/mbc.E20-08-0546} {\bibfield  {journal} {\bibinfo
  {journal} {Mol Biol Cell}\ }\textbf {\bibinfo {volume} {32}},\ \bibinfo
  {pages} {435} (\bibinfo {year} {2021})}\BibitemShut {NoStop}%
\bibitem [{\citenamefont {Berezhkovskii}\ \emph {et~al.}(1989)\citenamefont
  {Berezhkovskii}, \citenamefont {Makhnovskii},\ and\ \citenamefont
  {Suris}}]{berezhkovskii1989wiener}%
  \BibitemOpen
  \bibfield  {author} {\bibinfo {author} {\bibfnamefont {A.}~\bibnamefont
  {Berezhkovskii}}, \bibinfo {author} {\bibfnamefont {Y.~A.}\ \bibnamefont
  {Makhnovskii}},\ and\ \bibinfo {author} {\bibfnamefont {R.}~\bibnamefont
  {Suris}},\ }\bibfield  {title} {\bibinfo {title} {Wiener sausage volume
  moments},\ }\href@noop {} {\bibfield  {journal} {\bibinfo  {journal} {J Stat
  Phys}\ }\textbf {\bibinfo {volume} {57}},\ \bibinfo {pages} {333} (\bibinfo
  {year} {1989})}\BibitemShut {NoStop}%
\bibitem [{\citenamefont {Agrawal}\ \emph {et~al.}(2022)\citenamefont
  {Agrawal}, \citenamefont {Scott},\ and\ \citenamefont
  {Koslover}}]{agrawal2022morphology}%
  \BibitemOpen
  \bibfield  {author} {\bibinfo {author} {\bibfnamefont {A.}~\bibnamefont
  {Agrawal}}, \bibinfo {author} {\bibfnamefont {Z.~C.}\ \bibnamefont {Scott}},\
  and\ \bibinfo {author} {\bibfnamefont {E.~F.}\ \bibnamefont {Koslover}},\
  }\bibfield  {title} {\bibinfo {title} {Morphology and transport in eukaryotic
  cells},\ }\href@noop {} {\bibfield  {journal} {\bibinfo  {journal} {Ann Rev
  Biophys}\ }\textbf {\bibinfo {volume} {51}},\ \bibinfo {pages} {247}
  (\bibinfo {year} {2022})}\BibitemShut {NoStop}%
\bibitem [{\citenamefont {Kapitein}\ and\ \citenamefont
  {Hoogenraad}(2011)}]{kapitein2011way}%
  \BibitemOpen
  \bibfield  {author} {\bibinfo {author} {\bibfnamefont {L.~C.}\ \bibnamefont
  {Kapitein}}\ and\ \bibinfo {author} {\bibfnamefont {C.~C.}\ \bibnamefont
  {Hoogenraad}},\ }\bibfield  {title} {\bibinfo {title} {Which way to go?
  cytoskeletal organization and polarized transport in neurons},\ }\href@noop
  {} {\bibfield  {journal} {\bibinfo  {journal} {Mol Cell Neurosci}\ }\textbf
  {\bibinfo {volume} {46}},\ \bibinfo {pages} {9} (\bibinfo {year}
  {2011})}\BibitemShut {NoStop}%
\bibitem [{\citenamefont {Guedes-Dias}\ and\ \citenamefont
  {Holzbaur}(2019)}]{guedes2019axonal}%
  \BibitemOpen
  \bibfield  {author} {\bibinfo {author} {\bibfnamefont {P.}~\bibnamefont
  {Guedes-Dias}}\ and\ \bibinfo {author} {\bibfnamefont {E.~L.}\ \bibnamefont
  {Holzbaur}},\ }\bibfield  {title} {\bibinfo {title} {Axonal transport:
  Driving synaptic function},\ }\href@noop {} {\bibfield  {journal} {\bibinfo
  {journal} {Science}\ }\textbf {\bibinfo {volume} {366}},\ \bibinfo {pages}
  {eaaw9997} (\bibinfo {year} {2019})}\BibitemShut {NoStop}%
\bibitem [{\citenamefont {Sarpangala}\ \emph {et~al.}(2024)\citenamefont
  {Sarpangala}, \citenamefont {Randell}, \citenamefont {Gopinathan},\ and\
  \citenamefont {Kogan}}]{sarpangala2024tunable}%
  \BibitemOpen
  \bibfield  {author} {\bibinfo {author} {\bibfnamefont {N.}~\bibnamefont
  {Sarpangala}}, \bibinfo {author} {\bibfnamefont {B.}~\bibnamefont {Randell}},
  \bibinfo {author} {\bibfnamefont {A.}~\bibnamefont {Gopinathan}},\ and\
  \bibinfo {author} {\bibfnamefont {O.}~\bibnamefont {Kogan}},\ }\bibfield
  {title} {\bibinfo {title} {Tunable intracellular transport on converging
  microtubule morphologies},\ }\href@noop {} {\bibfield  {journal} {\bibinfo
  {journal} {Biophys Rep}\ }\textbf {\bibinfo {volume} {4}} (\bibinfo {year}
  {2024})}\BibitemShut {NoStop}%
\bibitem [{\citenamefont {Ester}\ \emph {et~al.}(1996)\citenamefont {Ester},
  \citenamefont {Kriegel}, \citenamefont {Sander},\ and\ \citenamefont
  {Xu}}]{Ester1996DBSCAN}%
  \BibitemOpen
  \bibfield  {author} {\bibinfo {author} {\bibfnamefont {M.}~\bibnamefont
  {Ester}}, \bibinfo {author} {\bibfnamefont {H.-P.}\ \bibnamefont {Kriegel}},
  \bibinfo {author} {\bibfnamefont {J.}~\bibnamefont {Sander}},\ and\ \bibinfo
  {author} {\bibfnamefont {X.}~\bibnamefont {Xu}},\ }\bibfield  {title}
  {\bibinfo {title} {A density-based algorithm for discovering clusters in
  large spatial databases with noise},\ }in\ \href
  {https://api.semanticscholar.org/CorpusID:355163} {\emph {\bibinfo
  {booktitle} {Knowledge Discovery and Data Mining}}}\ (\bibinfo {year}
  {1996})\BibitemShut {NoStop}%
\bibitem [{\citenamefont {Chen}\ \emph {et~al.}(2015)\citenamefont {Chen},
  \citenamefont {Wang},\ and\ \citenamefont {Granick}}]{chen2015memoryless}%
  \BibitemOpen
  \bibfield  {author} {\bibinfo {author} {\bibfnamefont {K.}~\bibnamefont
  {Chen}}, \bibinfo {author} {\bibfnamefont {B.}~\bibnamefont {Wang}},\ and\
  \bibinfo {author} {\bibfnamefont {S.}~\bibnamefont {Granick}},\ }\bibfield
  {title} {\bibinfo {title} {Memoryless self-reinforcing directionality in
  endosomal active transport within living cells},\ }\href@noop {} {\bibfield
  {journal} {\bibinfo  {journal} {Nat Mater}\ }\textbf {\bibinfo {volume}
  {14}},\ \bibinfo {pages} {589} (\bibinfo {year} {2015})}\BibitemShut
  {NoStop}%
\bibitem [{\citenamefont {Fedotov}\ \emph {et~al.}(2018)\citenamefont
  {Fedotov}, \citenamefont {Korabel}, \citenamefont {Waigh}, \citenamefont
  {Han},\ and\ \citenamefont {Allan}}]{fedotov2018memory}%
  \BibitemOpen
  \bibfield  {author} {\bibinfo {author} {\bibfnamefont {S.}~\bibnamefont
  {Fedotov}}, \bibinfo {author} {\bibfnamefont {N.}~\bibnamefont {Korabel}},
  \bibinfo {author} {\bibfnamefont {T.~A.}\ \bibnamefont {Waigh}}, \bibinfo
  {author} {\bibfnamefont {D.}~\bibnamefont {Han}},\ and\ \bibinfo {author}
  {\bibfnamefont {V.~J.}\ \bibnamefont {Allan}},\ }\bibfield  {title} {\bibinfo
  {title} {Memory effects and l{\'e}vy walk dynamics in intracellular transport
  of cargoes},\ }\href@noop {} {\bibfield  {journal} {\bibinfo  {journal} {Phys
  Rev E}\ }\textbf {\bibinfo {volume} {98}},\ \bibinfo {pages} {042136}
  (\bibinfo {year} {2018})}\BibitemShut {NoStop}%
\bibitem [{\citenamefont {Schulze}\ and\ \citenamefont
  {Kirschner}(1986)}]{schulze1986microtubule}%
  \BibitemOpen
  \bibfield  {author} {\bibinfo {author} {\bibfnamefont {E.}~\bibnamefont
  {Schulze}}\ and\ \bibinfo {author} {\bibfnamefont {M.}~\bibnamefont
  {Kirschner}},\ }\bibfield  {title} {\bibinfo {title} {Microtubule dynamics in
  interphase cells.},\ }\href@noop {} {\bibfield  {journal} {\bibinfo
  {journal} {J Cell Biol}\ }\textbf {\bibinfo {volume} {102}},\ \bibinfo
  {pages} {1020} (\bibinfo {year} {1986})}\BibitemShut {NoStop}%
\bibitem [{\citenamefont {Fritzsche}\ \emph {et~al.}(2013)\citenamefont
  {Fritzsche}, \citenamefont {Lewalle}, \citenamefont {Duke}, \citenamefont
  {Kruse},\ and\ \citenamefont {Charras}}]{fritzsche2013analysis}%
  \BibitemOpen
  \bibfield  {author} {\bibinfo {author} {\bibfnamefont {M.}~\bibnamefont
  {Fritzsche}}, \bibinfo {author} {\bibfnamefont {A.}~\bibnamefont {Lewalle}},
  \bibinfo {author} {\bibfnamefont {T.}~\bibnamefont {Duke}}, \bibinfo {author}
  {\bibfnamefont {K.}~\bibnamefont {Kruse}},\ and\ \bibinfo {author}
  {\bibfnamefont {G.}~\bibnamefont {Charras}},\ }\bibfield  {title} {\bibinfo
  {title} {Analysis of turnover dynamics of the submembranous actin cortex},\
  }\href@noop {} {\bibfield  {journal} {\bibinfo  {journal} {Mol Biol Cell}\
  }\textbf {\bibinfo {volume} {24}},\ \bibinfo {pages} {757} (\bibinfo {year}
  {2013})}\BibitemShut {NoStop}%
\end{thebibliography}%

\end{document}